\documentclass[american,twocolumn,aps,pre]{revtex4-1}
\usepackage[T1]{fontenc}
\usepackage[latin9]{inputenc}
\usepackage{verbatim}
\usepackage{graphicx}
\usepackage{esint}

\makeatletter

\providecommand{\tabularnewline}{\\}
\newcommand{\lyxdot}{.}

\@ifundefined{textcolor}{}
{%
 \definecolor{BLACK}{gray}{0}
 \definecolor{WHITE}{gray}{1}
 \definecolor{RED}{rgb}{1,0,0}
 \definecolor{GREEN}{rgb}{0,1,0}
 \definecolor{BLUE}{rgb}{0,0,1}
 \definecolor{CYAN}{cmyk}{1,0,0,0}
 \definecolor{MAGENTA}{cmyk}{0,1,0,0}
 \definecolor{YELLOW}{cmyk}{0,0,1,0}
}

\makeatother

\usepackage{babel}
\begin{document}

\title{Mean properties and Free Energy of a few hard spheres confined in
a spherical cavity}

\author{I. Urrutia$^{\dag*}$, C. Pastorino$^{\dag*}$}
\begin{abstract}
We use analytical calculations and event-driven molecular dynamics
simulations to study a small number of hard sphere particles in a
spherical cavity. The cavity is taken also as the thermal bath so
that the system thermalizes by collisions with the wall. In that way,
these systems of two, three and four particles, are considered in
the canonical ensemble. We characterize various mean and thermal properties
for a wide range of number densities. We study the density profiles,
the components of the local pressure tensor, the interface tension,
and the adsorption at the wall. This spans from the ideal gas limit
at low densities to the high-packing limit in which there are significant
regions of the cavity for which the particles have no access, due
the conjunction of excluded volume and confinement. The contact density
and the pressure on the wall are obtained by simulations and compared
to exact analytical results. We also obtain the excess free energy
for $N=4$, by using a \emph{simulated-assisted} approach in which
we combine simulation results with the knowledge of the exact partition
function for two and three particles in a spherical cavity.
\end{abstract}

\affiliation{$^{*}$Departamento de Física de la Materia Condensada, Centro Atómico
Constituyentes, CNEA, Av.Gral.~Paz 1499, 1650 Pcia.~de Buenos Aires,
Argentina}

\affiliation{$^{\dag}$ CONICET, Avenida Rivadavia 1917, C1033AAJ Buenos Aires,
Argentina}

\email{iurrutia@cnea.gov.ar}

\email{pastor@cnea.gov.ar}

\maketitle

\section{Introduction}

The properties of a few particles confined in a spherical cavity has
become attractive recently, due to the possibility of synthesizing
the so-called silica nano-rattles\cite{Tan_2010} and their promising
wide range of applications. An important biomedical application of
these mesoporous silica nanoparticles is in the area of drug delivery,
targeted to disease diagnosis and therapy\cite{Tang_12}. Within the
field of core-shell systems, hollow silica nano-particles containing
smaller particles or molecules were synthesized in different conditions
and packed as mesoporous materials\cite{Chen_10,Tang_12}. Single
silica nano-containers confining gold nano-particles have been proposed
as an option for drug delivery vehicles in cancer therapies\cite{Yang_08}
and also as imaging probes with fluorescent dies\cite{Wang_11}. Hollow
nanospheres with mesoporous shells and metal cores have also interesting
optical and electrical properties, which are relevant in imaging and
catalysis applications. Hindering the interaction between metal nanoparticles,
by encapsulating them by mesoporous shells, keep the metallic particles
active for longer times. These systems are known as yolk-shell nanoparticles
and are also promising as nanoreactors, in bio-medicine applications
and photo-catalysis\cite{Tan_2010,Liu_10}.

From a theoretical point of view, a few particles confined in a cavity
can be regarded as the few-bodies limit of the liquid or fluid states
and also as a limiting case in which the curvature of the confining
wall is very significant. Systems of many particles interacting through
simple potentials, that form bulk fluid phases, confined in spherical
cavities produce a rich set of structural and thermal properties and
were studied thoroughly \cite{Dinsmore_1998,Statt_2012,Huang_2013}.

Particles interacting through a hard sphere (HS) potential are widely
used as models for colloids and simple fluids\cite{Roth_2010,Zykova_2010,Winkler_2013}.
The HS model was used in pioneering works of molecular dynamics simulations
to study phase transitions and elastic media\cite{Alder_1957,Alder_1963,Alder_1968}.

The spherically confined HS system has been studied using molecular
dynamics (MD)\cite{Macpherson_1987}, Monte Carlo simulations (MC)
and density functional theories (DFT)\cite{Gonzalez_1997,Gonzalez_2006},
mainly in the context of liquid state theory. However, the limit of
few particles confined in a spherical cavity has received much less
attention. One of us performed analytical calculations for two and
three HS confined in a spherical cavity\cite{Urrutia_2008,Urrutia_2010,Urrutia_2010b,Urrutia_2011_b}
and in pores of different shapes\cite{Urrutia_2010b}. Huang and co-workers
studied recently, by molecular dynamics simulation the structure and
equation of state of hard sphere confined in a small spherical geometry,
with a focus in a freezing-like transition at high densities \cite{Huang_2013}.
Based on the HS interaction, Winkler et al. analyzed phase transitions
of a colloid-polymer mixture inside a hard spherical cavity \cite{Winkler_2013}.
New insights on the curvature dependence of the surface free energy
of a fluid in contact with a curved wall were also revealed, by studying
the HS fluid \cite{Blokhuis_2013,Urrutia_2014}.

\begin{figure}[t]
\begin{centering}
\includegraphics[clip,width=0.8\columnwidth]{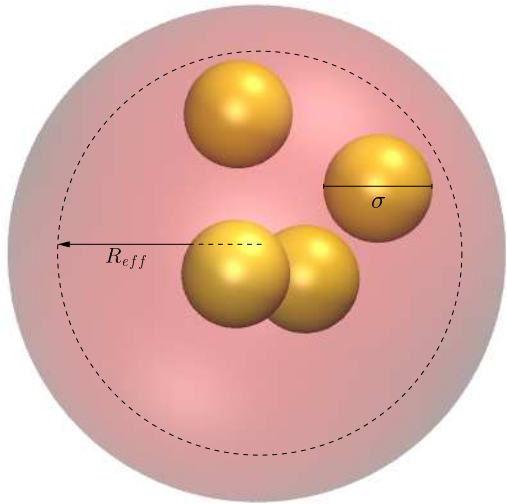}
\par\end{centering}

\protect\caption{(color online) A configuration of four hard spheres (yellow) confined
in a spherical cavity with a number density $\rho=0.5\sigma^{-3}$,
as obtained from an event-driven molecular dynamics simulation. The
sphere of radius $R_{o}=1.74\sigma$, is shown with transparency.
The diameter $\sigma$ of the HS particle is indicated in one of them,
and the effective radius $R_{\textrm{eff}}\equiv R_{o}-\frac{\sigma}{2}$
of the cavity is presented in dashed line. The snapshot was generated
with the VMD program\cite{VMD}.\label{fig:4hs}}
\end{figure}

In this work, we study the free energy and the structural and mean
properties of few-body HS confined in a spherical cavity, as a function
of the number of particles. For the sake of clarity we give operational
expressions for configuration integral $Z_{4}$ and the pressure on
the wall $P_{w}$ as a function of the effective cavity radius. Two
different approaches are used: we perform event-driven molecular dynamics\cite{Allen_and_Tildesley}
with a fixed-temperature spherical wall and compare the results with
exact or quasi-exact theoretical calculations, obtained from analytical
expressions of the free energy. Our minimal model utilizes the simple
HS interaction potential between particles and a spherical hard boundary,
chosen to capture the purely geometric effect of the spherical confinement
and the behavior of the translational entropy of the particles, for
a wide range of number densities. The number of particles $N$ and
$T$ are kept fixed. A typical configuration of the system with $N=4$
particles, as obtained from computer simulations, and the relevant
length scales of the system are shown in Fig. \ref{fig:4hs}.

As the statistical mechanical ensembles equivalence does not apply
for few-particle systems, the approach is based on the canonical ensemble
partition function (CPF). We go through the theoretical details in
section \ref{sec:Theory} and devote section \ref{sec:Simulation-techniques}
to describe the simulation techniques and details. The combination
of analytical and simulation results to obtain the expressions for
$Z_{4}$ and the free energy is described in section \ref{sec:Simulation-assisted-quasi-exact-}.
The mean properties of the system and the free energy for 2, 3 and
4 HS particles in a spherical cavity are given in section \ref{results}.
Final remarks and conclusions are presented in section \ref{sec:Conclusions}.

\section{Theory\label{sec:Theory}}

The system under study is a set of $N$ HS-particles with hard repulsion
distance $\sigma$, confined in a spherical cavity with radius $R_{o}$.
The particles interact with the cavity through a hard wall potential
which prevents that particles escape to the outside. Thus, the effective
radius of the cavity is $R_{\textrm{eff}}\equiv R_{o}-\sigma/2$,
which represents the maximum possible distance between the center
of the cavity and the center of each HS. The temperature of the fluid
is determined by the wall temperature $T$, which is fixed. 

The statistical mechanical and thermodynamic properties of the system
are obtained from its CPF, $Q_{N}$. One actually works with the configuration
integral (CI), given that kinetic degrees of freedom integrate trivially.
Thus, the CPF reads
\begin{equation}
Q_{N}=\frac{1}{N!}\Lambda^{-3N}Z_{N}\:.\label{eq:CPF}
\end{equation}
Here $\Lambda=h/(2\pi m\, k_{B}T)^{1/2}$ is the thermal de Broglie
wavelength, while $m$ is the mass of each particle, $h$ is the Planck's
constant and $k_{B}$ is the Boltzmann constant. In addition, $Z_{N}$
is the CI of the system that accounts for its non-trivial properties
\begin{equation}
Z_{N}=\dotsint_{sph}\prod_{<j,k>}e_{jk}\, d^{N}\mathbf{r}\:,\label{eq:ZN}
\end{equation}
here $e_{jk}=\exp\left[-\beta\phi\left(r_{jk}\right)\right]=\Theta\left(r_{jk}-\sigma\right)$
is the Boltzmann factor, $\phi$ is the HS pair potential, $r_{jk}$
the distance between the $j$ and $k$ particles, and the Heaviside
function is $\Theta\left(x\right)=1$ if $x\geq0$ and $\Theta\left(x\right)=0$
otherwise. The inverse temperature is defined as $\beta=1/T$ (for
simplicity we fix $k_{B}\equiv1$). The spatial domain of the integral
for each particle is constrained to the spherical cavity. Naturally,
$Z_{N}$ is a function of the cavity's radius $R_{o}$ and $\sigma$.
The CI for $N=1$ is identified with the volume of the system through
$Z_{1}=V$ ($V=4\pi R_{\textrm{eff}}^{3}/3$), and the next two terms
$Z_{2}$ and $Z_{3}$ were evaluated and discussed in Ref. \cite{Urrutia_2008,Urrutia_2011_b}.
In general, the structure of $Z_{N}$ is related to the $Z_{i}$'s
with $i<N$ through a well known polynomial form (see p.135 in Ref.
\cite{Hill1956}), which also involves the inhomogeneous system cluster
integral of $N$ particles, $\tau_{N}$, as constant term. Alternatively,
$Z_{N}$ can also be expressed as a polynomial in $\tau_{i}$'s with
$i\leq N$. $\tau_{N}$ was introduced by Mayer in his cluster theory
of homogeneous system with a different notation, and was latter used
to develop the diluted gas equation of state (EOS), known as Virial
series. A general form for the cluster integrals of the HS system
confined in a spherical cavity is\cite{Urrutia_2011_b,Urrutia_2012}
\begin{equation}
\tau_{N}/N!=V\, b_{N}-\, A\, a_{N}+\mathsf{J}\, c_{N,\mathsf{J}}+\mathsf{K}\, c_{N,\mathsf{K}}+\mathsf{S}_{N}(R_{\textrm{eff}}^{-1})+\Delta\tau_{N}/N!\;,\label{eq:biVgen}
\end{equation}
where $A=4\pi R_{\textrm{eff}}^{2}$, $\mathsf{J}=8\pi R_{\textrm{eff}}$,
$\mathsf{K}=4\pi$ are the surface area and the mean and Gaussian
extensive curvatures, respectively. $\mathsf{S}_{N}$ includes the
higher order terms in $R_{\textrm{eff}}^{-1}$, while $\Delta\tau_{N}$
is the non-analytic core, being $\Delta\tau_{N}=0$ for $R_{\textrm{eff}}>\left(N-1\right)\sigma$.
The constant coefficients $\left\{ b_{N},a_{N},c_{N,\mathsf{J}},c_{N,\mathsf{K}}\right\} $
are the volume, area and curvature components of $\tau_{N}$. The
known components up to order four are presented in Table \ref{tab:TauiComp}.
For $N=2,\,3$ and $4$ the CIs are \cite{Yang_2013} 
\begin{eqnarray}
Z_{2} & = & Z_{1}^{2}+\tau_{2}\:,\nonumber \\
Z_{3} & = & -2Z_{1}^{3}+3Z_{1}Z_{2}+\tau_{3}\:,\nonumber \\
Z_{4} & = & 6Z_{1}^{4}-12Z_{1}^{2}Z_{2}+3Z_{2}^{2}+4Z_{1}Z_{3}+\tau_{4}\:.\label{eq:Z234}
\end{eqnarray}
\begin{table}[t]
\begin{centering}
\begin{tabular}{|c|c|c|c|c|c|}
\hline 
$N$ & $b_{N}$ & $a_{N}$ & $c_{N,\mathsf{J}}$ & $c_{N,\mathsf{K}}$ & $\mathsf{S}_{N}$, $\Delta\tau_{N}$\tabularnewline
\hline 
\hline 
$2$ & $-2.09439$ & $-0.39270$ & $0$ & $-0.02182$ & $0$ \tabularnewline
\hline 
$3$ & $7.40220$ & $2.4145$ & $0.13469$ & $0.21242$ & $\neq0$\tabularnewline
\hline 
$4$ & $-32.6506$ & $-14.3871(6)$ & ${\rm unknown}$ & ${\rm unknown}$ & ${\rm unknown}$\tabularnewline
\hline 
\end{tabular}
\par\end{centering}

\protect\caption{The volume, surface area and curvatures components of $\tau_{N}$
for a HS system (here we set $\sigma=1$) \cite{Urrutia_2014}. See
Eq. (\ref{eq:biVgen}).\label{tab:TauiComp}}
\end{table}
 Two limiting regimes characterize the features of $Z_{N}$ independently
of the number of particles in the cavity. For very large values of
$R_{\textrm{eff}}$ (very small number densities) the CI scales as
$Z_{N}\simeq V^{N}$ and the system behaves basically as an ideal
gas. In the opposite limit, for each $N$ there exist a small enough
cutoff value of $R_{\textrm{eff}}$, the packing radius $R_{\textrm{m}}$,
such that $R_{\textrm{eff}}<R_{\textrm{m}}$ implies that the hard
spheres do not fit in the cavity. Thus $Z_{N}=0$. Close to this high-density
limit, at $R_{\textrm{eff}}\gtrsim R_{\textrm{m}}$, the behavior
of $Z_{N}$ is characterized by its root, while $\tau_{N}$ is finite
and the pressure on the wall diverges as will be shown in the next
section.

\subsection{Free energy of a few body system}

The standard relations between CPF and thermal quantities, that are
usually applied to systems with a large number of particles, also
apply to few-body systems\cite{Hansen2006}. Of course, this statement
is valid provided that any assumption about the large number of particles
involved and the so-called thermodynamic limit are avoided\cite{Urrutia_2010b,Urrutia_2012}.
Thus, for a system in stationary conditions with fixed $N$ and $T$
the Helmholtz's free energy ($F$) is 
\begin{equation}
\beta F=-\ln Q\:,\label{eq:FlogQ}
\end{equation}
\begin{equation}
\beta F=\beta U-S/k_{B}\:,\label{eq:1rstLaw}
\end{equation}
where $U$ is the system energy and $S$, its entropy. The energy
term of HS particles is simply $\beta U=3N/2$. The reversible work
done at constant temperature, to change the cavity radius between
states $a$ and $b$ is equal to (minus) the change of free energy
\begin{equation}
F_{b}-F_{a}=-\intop_{a}^{b}P_{w}dV\:,\label{eq:DiffF}
\end{equation}
with $dV=A\, dR_{\textrm{eff}}$. Here, $P_{w}$ is the pressure on
the spherical wall which is an EOS of the system. The derivative
of Eq. (\ref{eq:DiffF}) at constant $T$ gives 
\begin{equation}
P_{w}A=-dF/dR_{\textrm{eff}}\:,\label{eq:PwThermo}
\end{equation}
which meets the exact relation known as contact theorem\cite{Hansen2006,Blokhuis_2007},
\begin{equation}
\beta P_{w}=\rho\left(R_{\textrm{eff}}\right)\:.\label{eq:rhoscontact}
\end{equation}
In this ideal gas-like relation, $\rho\left(R_{\textrm{eff}}\right)$
is the value that takes the density profile at contact with the wall.
This is an extended version of the contact theorem for planar walls.
In this case, it applies to curved walls of constant curvature (spheres
and cylinders), for both open and closed systems. Note that Eq. (\ref{eq:rhoscontact})
follows directly from Eqs. (\ref{eq:CPF},\ref{eq:ZN},\ref{eq:FlogQ},\ref{eq:DiffF})
and (\ref{eq:PwThermo}), where in Eq. (\ref{eq:ZN}) it is convenient
to explicitly write the Boltzmann factors for the external potential,
i.e. $\Theta(R_{\textrm{eff}}-r_{i})$.

As it was mentioned above, the use of Eq. (\ref{eq:CPF}) to Eq. (\ref{eq:rhoscontact})
is well established for many-body systems. Yet, these Eqs. also apply
to the much less studied case of few-body systems. As this might be
regarded as controversial, we provide here a few arguments to support
it \cite{Munakata_2002,Urrutia_2010b,Urrutia_2012}. An important
point is that statistical mechanics theory of ensembles does not exclude
few-body systems%
\footnote{See for example p. 134 in the book by Hill \cite{Hill1956} or p.
28 in the book of Hansen and McDonald\cite{Hansen2006}%
}. In addition, independently of the considered number of particles,
we relate the system properties with the corresponding mean-ensemble
and thermal quantities such as energy, free energy, entropy. Thus,
we follow the basic assumptions of statistical mechanics for many-body
systems, except those that explicitly consider a large number of particles,
to postulate the validity of Eqs. (\ref{eq:FlogQ}) and (\ref{eq:1rstLaw})
for few-body systems. In particular, we avoid the ensembles equivalence,
that is valid for large number of particles but does not apply to
few-body systems. Finally, the ergodicity of the system is also assumed
to claim the equivalence between the mean ensemble value of a magnitude
and the simulation time average. Ultimately, the comparison of theoretical
results with simulation or experimental outcome serves to validate
our approach.

\section{Simulation techniques\label{sec:Simulation-techniques}}

As it was already mentioned, in few-body systems, the statistical
mechanical equivalence between different ensembles does not apply.
Therefore, simulation and statistical mechanical approches should
correspond in detail to the same physical conditions, to ensure that
the obtained results are comparable. In this work, we focus on a system
at constant temperature. From the statistical mechanical theory point
of view, it corresponds to a canonical ensemble, and thus one assumes
a Maxwell-Boltzmann velocity distribution. On the other hand, the
simulations should include a thermostating mechanism that ensures
constant temperature and Maxwell-Boltzmann velocity distribution.
The effect of utilizing different ensembles on the properties of few,
spherically confined, hard-spheres was analyzed in Ref. \cite{Gonzalez_1998}.

Molecular dynamics simulations of few hard spheres confined in a spherical
cavity were performed with a standard event-driven algorithm (EDMD)\cite{Allen_and_Tildesley}.
Constant temperature was achieved by using a thermal walls-thermostat,
which changes the velocity of the particle colliding with the wall
by means of a velocity distribution compatible with canonical ensemble
for a given temperature $T$. We emphasize that this way of thermostating
the system is unusual for bulk simulations. For these, local thermostats
acting on each particle, like the Langevin thermostat, or a global
thermostat that act on the system as a whole, by means of extra degrees
of freedom and dynamical equations (i.e. Nosé-Hoover) are usually
employed\cite{Hunenberger_05}. In our case, the particles are thermalized
by collisions with the wall, in the same way in which a real system
exchange heat with a confining medium, which acts also as thermal
bath\cite{Yong_2013}. We consider that this procedure is very representative
of the experimental case and, as we deal with very small systems,
the collisions with the wall are very frequent which produce an efficient
thermalization. We notice also, that the HS-system is usually considered
athermal because the interacting potential does not introduce an energy
scale%
. For example, the isotherms $P(T)$ vs $\rho$ collapse in a unique
curve when plotted $\beta P$ vs. $\rho$. In the simulation, however
we consider the HS system in contact with a thermal bath which imposes
its temperature to the system. In this line of thinking, we consider
that the HS-system is in thermal equilibrium at a fixed temperature. 

In this way, we account for two types of events, namely, particle-particle
collision and particle-wall collision. For the particle-particle collision,
the usual EDMD algorithm was used, in which the particle moves with
rectilinear and constant-velocity dynamics, between particle collisions\cite{Allen_and_Tildesley}.

\begin{figure}[t]
\begin{centering}
\includegraphics[width=0.98\columnwidth]{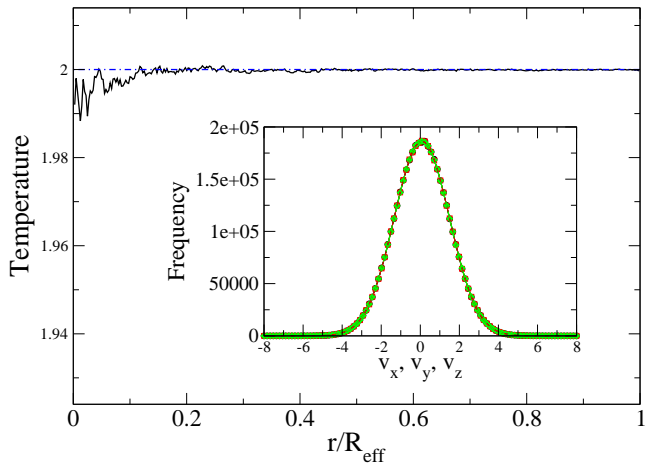}
\par\end{centering}

\protect\caption{(color online) Temperature profile across a spherical cavity with
three particles and number density of $\rho=0.1$ for a canonical
simulation with thermal walls thermostat. The temperature is set to
$T=2$. The temperature obtained from the kinetic energy of the particles
follows this value with very good accuracy in most of the cavity.
In the center of the cavity, higher fluctuations and deviations from
the imposed value are observed, due to the poorer statistics in this
region. This is a feature taking place for all the profiles measured
across the cavity. The inset shows the velocity distribution functions
for $v_{x}$,$v_{y}$ and $v_{z}$(symbols, strongly overlapped).
An excellent gaussian fit, which corroborates the Maxwell-Boltzmann
distribution is also shown in dashed line. The variance of the distribution
for $v_{x}$ for example, is $\sigma^{2}\equiv\frac{k_{B}T}{m}=2.0012$,
giving a very good value for the temperature ($m=1$). This behavior
was verified for different number densities, number of particles and
temperatures. \label{fig:temp_prof}}
\end{figure}

The time to collision of a particle $i$ approaching a particle $j$
is calculated as:
\begin{equation}
t_{ij}=\frac{-b_{ij}-\left(b_{ij}^{2}-v_{ij}^{2}(r_{ij}^{2}-\sigma^{2})\right)^{1/2}}{v_{ij}^{2}}\,,\label{eq:t_ij_edmd}
\end{equation}
where $\mathbf{r}_{ij}\equiv\mathbf{r}_{i}-\mathbf{r}_{j}$ and $\mathbf{v}_{ij}\equiv\mathbf{v}_{i}-\mathbf{v}_{j}$
are the relative positions and velocities of the particle pair, respectively.
The parameter $b_{ij}\equiv\mathbf{r}_{ij}\cdot\mathbf{v}_{ij}$ must
be negative if the particles are approaching each other and positive
otherwise, and we consider only the positive values of $t_{ij}$.
Here, it is assumed that the spheres are not already overlapping.
The Eq. (\ref{eq:t_ij_edmd}) is obtained by simply asking that at
collision time $t_{ij}$, the condition $|\mathbf{r}_{ij}+\mathbf{v}_{ij}t_{ij}|=\sigma$
must be fulfilled. An ordered list of events, with increasing collision
times $t_{ij}$ is generated. Between collisions times, the particles
move with $\mathbf{r}_{i}=\mathbf{v}_{i}t$. Once a collision occurs,
the new velocities of the pair of particles involved in the collision
are obtained as:
\begin{eqnarray*}
\mathbf{v}_{i}^{{\rm new}} & = & \mathbf{v}_{i}^{{\rm old}}+\delta\mathbf{v}\\
\mathbf{v}_{j}^{{\rm new}} & = & \mathbf{v}_{j}^{{\rm old}}-\delta\mathbf{v}\,\,,
\end{eqnarray*}
where $\delta\mathbf{v}=-(b_{ij}/\sigma^{2})\mathbf{r}_{ij}$\cite{Allen_and_Tildesley}.
In our case, however, it must be also considered the time for which
each particle collides with the wall $t_{i}^{w}$. This time is calculated
by the condition
\[
|\mathbf{r}_{i}+\mathbf{v}_{i}t_{i}^{w}|=R_{o}-\frac{\sigma}{2}\,,
\]
The nearest next event is chosen as the minimum of the next particle
and wall events: $\min(\min(t_{ij}),\min(t_{i}^{w}))$. If the particle-wall
collision is the next event, the system is evolved until the particle
reaches the wall. At this point, the thermal-walls thermostat acts
on the particle by imposing it a new velocity, which is chosen stochastically
from the probability distribution densities:
\begin{eqnarray}
p_{n}(v_{n}) & = & m\beta|v_{n}|\exp\left(-\beta\frac{1}{2}mv_{n}^{2}\right)\label{eq:thermal_wall}\\
p_{t}(v_{t}) & = & \sqrt{\frac{m\beta}{2\pi}}\exp\left(-\beta\frac{1}{2}mv_{t}^{2}\right)\:,\nonumber 
\end{eqnarray}
here $n$ and $t$ stands for the normal and tangential components
of the velocities, that lie in directions $-\hat{r}$ and $\mathbf{v}^{{\rm old}}-(\mathbf{v}^{{\rm old}}\cdot\hat{r})\hat{r}$,
respectively. The thermal walls described by Eq. (\ref{eq:thermal_wall})
fix the temperature of the system. They were tested for HS confined
by planar walls and produce a velocity distribution compatible with
that of Maxwell-Boltzmann\cite{Tehver_98}. In the present study we
have verified the same behavior for the temperature profile and the
distribution of the velocities for curved walls. It should be noted
that it is not correct to simply use a Gaussian distribution for the
normal component of the velocity bouncing in the wall (Eq. (\ref{eq:thermal_wall})).
This can be rationalized by understanding that the flux of particles
leaving the wall is what must be sampled and not the distribution
of velocities of particles in the wall\cite{Tehver_98,Inoue_02}.
In Fig. \ref{fig:temp_prof} we provide an example of temperature
profile, for a number density $\rho=0.1$, $N=3$ and $T=2$. The
local temperature across the cavity agrees with the imposed value,
and higher fluctuations and deviations are observed only close to
the center of the cavity, due to poorer statistics. This feature is
typical of all the profiles of local quantities for the spherical
geometry. The Inset shows the distribution of velocities of the particles,
giving an excellent fit to a Gaussian distribution of $T=2$ (see
caption). This behavior was verified for all the densities and number
of particles considered in this work. All the simulation results presented
throughout this work correspond to $T=2$.

Different number of particles were simulated using typically $4\times10^{7}$
particle-particle colliding events for each physical condition. The
densities were varied in a broad range from very low densities to
high densities, in the vicinity of the closed-packing condition.

During the simulation we evaluate various thermal quantities and spherically
averaged functions of position. The studied structure and strain position
functions are: the one body distribution function $\rho\left(\mathbf{r}\right)$
and the two independent components of the local pressure tensor: the
normal $P_{n}\left(\mathbf{r}\right)$ and tangent $P_{t}\left(\mathbf{r}\right)$
components, respectively. Naturally, the calculated profiles for $\rho\left(r\right)$,
$P_{t}\left(r\right)$ and $P_{n}\left(r\right)$ are mean values
over a discrete domain, obtained on a grid of spherical shells during
the elapsed simulation time. The grid takes steps of $\Delta r=R_{\textrm{eff}}/200$.
The definition for the local pressure tensor is that of Irving and
Kirkwood (IK), i.e. a straight line path for the contour joining two
centers of force \cite{Irving_1950,Schofield_1982}. The IK pressure
tensor is well behaved in spherical coordinates and can be written
as\cite{Hafskjold_2002}

\begin{equation}
P_{\alpha\beta}\left(\mathbf{r}\right)=\rho\left(\mathbf{r}\right)k_{B}T\delta_{\alpha\beta}+\frac{1}{2}\left\langle \sum_{i}\sum_{j\neq i}P_{ij}^{\alpha\beta}\left(\mathbf{r}\right)\right\rangle \:,\label{eq:Pn}
\end{equation}
where $\delta_{\alpha\beta}$ is the Kronecker delta and $\alpha,\beta$
run over the local orthogonal unit vectors in spherical coordinates
$\hat{r}$, $\hat{\theta}$ and $\hat{\phi}$. The configurational
contribution to the tensor involves the ensemble average $\left\langle \cdots\right\rangle $
and each pair of particles contributes with 
\begin{equation}
P_{ij}^{\alpha\beta}\left(\mathbf{r}\right)=-u'\left(r_{ij}\right)\intop_{0}^{1}\frac{r_{ij}^{\alpha}r_{ij}^{\beta}}{r_{ij}}\delta\left(\mathbf{r}-\mathbf{r}_{i}+\lambda\mathbf{r}_{ij}\right)d\lambda\:,\label{eq:Pijmicro}
\end{equation}
where $\delta\left(x\right)$ is the Dirac delta function. During
the simulation of the HS system, each collision contributes to the
pressure with
\begin{equation}
\Pi=m\sigma\frac{\delta v}{\delta t}\:.\label{eq:PIpressure}
\end{equation}
This contribution is distributed in $M$ points located at $\mathbf{r}_{l}$
($l=1,\cdots,\, M$), that are equally spaced along the line of length
$\sigma$, between the pair of colliding particles. Each point $\mathbf{r}_{l}$,
that lies in the $k$-th spherical slab of the grid with mean radius,
contributes to $P_{\alpha\beta}\left(r_{k}\right)$ with $\Pi\hat{r}_{ij}^{\alpha}\hat{r}_{ij}^{\beta}/M$
where the projections on the spherical coordinates are $\hat{r}_{ij}^{\alpha}=\hat{r}_{ij}.\hat{\alpha}\left(\mathbf{r}_{l}\right)$.
The two independent components of the diagonal tensor are $P_{n}=P_{rr}$
and $P_{t}=\left(P_{\theta\theta}+P_{\phi\phi}\right)/2$.%
{} The discretized domain produces average local values that can be
slightly different from the exact ones. Particularly, in the near-wall
and central regions, the binning of different volume produce a poorer
statistics for smaller $r$, which can be noticed in the profiles.

\section{\label{sec:Simulation-assisted-quasi-exact-}Simulation-assisted
quasi-exact analytical approach}

The CI of the system of four HS confined in a spherical cavity $Z_{4}(R_{\textrm{eff}})$
reduces to an integral in nine variables, as it is also the case of
the fourth cluster integral $\tau_{4}(R_{\textrm{eff}})$ of the system.
A taste of the difficulty of solving these integrals can be gained
by recognizing that, for the case of homogeneous HS system where $\tau_{N}=N!b_{N}V$,
the highest order cluster integral exactly known is the fourth (the
constant $b_{4}$ reduces to an integral in six variables).\cite{Nairn_1972}
On the other hand, the higher order cluster integral known for an
inhomogeneous HS system is $\tau_{3}(R_{\textrm{eff}})$ for spherical
confinement. This integral also reduces to a six-variable integral
\cite{Urrutia_2011_b}. In this work, we develop a \emph{simulation-assisted}
procedure that uses the mean simulation value of wall pressure of
the 4-HS in a hard-wall spherical cavity to obtain, based on fitting,
an analytic expressions for both $\tau_{4}$ and $Z_{4}$. These expressions
accurately describe both functions along the complete range of cavity
sizes $R_{\textrm{eff}}>0.6124$. In this section, we set $\sigma\equiv1$
to deal with simpler notation.

Cluster integrals of HS system in a spherical confinement are non
analytic functions of $R_{\textrm{eff}}.$ If we analyze from the
infinitely diluted configuration by decreasing the cavity radius,
the first analytic branch of $\tau_{N}$ extends from to $R_{\textrm{eff}}\rightarrow\infty$
to $R_{\textrm{eff}}=\left(N-1\right)/2.$ There, a discontinuity
in a higher order derivative may appear. For example, at $R_{\textrm{eff}}=1$
$\tau_{3}\left(R_{\textrm{eff}}\right)$ has a discontinuous fifth
order derivative.\cite{Urrutia_2011_b} For even smaller $R_{\textrm{eff}}$,
a sequence of non-analytic points appear in $\tau_{N}$ until the
highest density packing is attained at $R_{\textrm{eff}}=R_{\textrm{m}}$,
where $Z_{N}$ collapses to zero and $\tau_{N}$ takes a finite value.
For each $N$, the number density $\rho_{0}=N/V$%
{} reaches a different maximum driven by the geometry of the packing
configuration, that exceeds the packing density of the bulk system
$\rho_{0}=1.4142$. The system with $N=2$ attains its maximum density
$\rho_{0}=3.8197$ at $R_{\textrm{m}}=0.5$, for $N=3$ it is $\rho_{0}=3.7215$
at $R_{\textrm{m}}=0.5773$, while for $N=4$ it corresponds to $\rho_{0}=4.1584$
at $R_{\textrm{m}}=0.6124$.

By using the analytical expressions (\ref{eq:Z234}), we obtained
$Z_{4}$ by fitting simulation results of $P_{w}$ in the low and
high density branches in a process that we call simulation-assisted
calculation. A detailed explanation of this procedure is provided
in Appendix \ref{sec:Appendix}. 

A peculiarity of the $N=4$ system is that an ergodicity break is
produced when the cavity radius decreases below $R_{\textrm{eff}}=0.7071,$
that corresponds to $\rho_{0}=2.70095$. This purely classical effect,
that is absent for $N=2$ and $3$, appears for distinguishable particles,
but is absent for indistinguishable ones. In the ergodicity break,
the available region of the spatial domain of the particles separates
into two identical disjointed regions. This produces a discontinuous
step in the CI that decreases by a factor $1/2$. This curious behavior
appears because, for $R_{\textrm{eff}}<0.7071$, it is impossible
the exchange of positions between two particles, keeping the other
two fixed. The spatial constraint forces the 4 HS to move as it were
a tetrahedron, that can only rotate and have small deformations. Alternatively,
once the labeled particles are constrained to such a small cavity,
there is no path that allows to go from a given available state to
another one, obtained from its speculate reflection. Systems of 2
HS confined in cylindrical and cuboidal pores have also shown an ergodicity
break behavior \cite{Urrutia_2010b}.%

\section{\label{results}Mean properties and Free Energy of cavities with
$2,\,3$ and $4$ particles }

Even at this very small number of particles in the studied system,
we define a bulk-like behavior to compare the mean and thermal quantities
of the system, with their macroscopic homogeneous fluid counterparts.
These bulk-like behavior is found typically at intermediate densities
in the center of the cavity, where an approximate constant density
profile is observed. In many properties, we found a qualitatively
different behavior, depending upon the number of particles, whose
root is a different (geometrical) packing with varying number of particles.
We will discuss these differences while presenting the results.

\begin{figure}[t]
\begin{centering}
\includegraphics[clip,width=0.98\columnwidth]{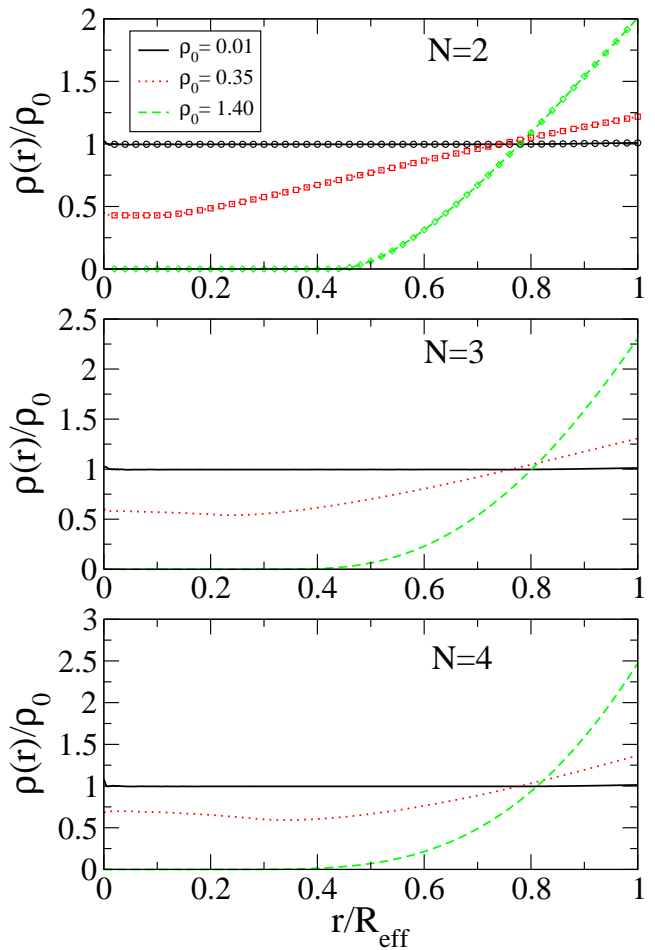}
\par\end{centering}

\protect\caption{(color online) Density profiles $\rho(r)$ across the spherical cavity
for three representative low (complete line), medium (point line)
and high (dashed line) number densities. From top to bottom panels
display results for $N=2,\,3$ and $4$ hard spheres particles, respectively.
For the case $N=2$, the exact analytical results are shown with open
symbols. The axis are scaled with the cavity radius to compare with
the same scale different number of particles.\label{fig:Density-profiles_low_high}}
\end{figure}
Figure \ref{fig:Density-profiles_low_high} shows the density profiles
across the spherical cavity from the center ($r=0$) towards the wall
of the cavity for three distinctive total number densities $\rho_{0}$.
The density profiles are scaled with the number density of the system
$\rho_{0}$ and the $r$ coordinate, with the effective radius of
the cavity $R_{\textrm{eff}}$ to allow comparison between different
number of particles. For 2, 3 and 4 particles, the overall features
are similar. At low number density ($\rho_{0}=0.01$), the profile
is constant (homogeneous) across the pore, indicating ideal gas behavior.
At intermediate densities, represented here by $\rho_{0}=0.35$, a
clear homogeneity in the density profile shows-up, with an increase
of density close to the wall. A central region, which we call plateau,
of approximately constant density is still present, but its extension
decreases towards the center of the cavity for higher number of particles.
At high number densities ($\rho_{0}=1.40\sigma^{-3}$) the inhomogeneous
profile persists, with the distinctive feature of vanishing density
in the central region. In this high packing regime, the particles
cannot occupy the center of the cavity, due to the presence of the
other ones. In the case of $N=2$, a direct comparison between theory
(open circles) and simulation for $\rho(r)/\rho_{0}$ along the entire
range of $r$ is presented. The agreement between simulation and
theory is excellent. This strengthens the confidence in the thermostating
scheme used in simulations and the overall theoretical and numerical
approaches. We also checked temperature conservation with the temperature
profiles (see Fig. \ref{fig:temp_prof}) across the cavity, obtained
by the application of the equipartition theorem to the local kinetic
energy of the particles. In the case of $N=3$, only $\rho_{\textrm{center}}$
and $\rho(R_{\textrm{eff}})$ were found theoretically. For $N=4$
the simulation-assisted approach and Eq. (\ref{eq:rhoscontact}) enables
to obtain $\rho(R_{\textrm{eff}})$. 

\begin{figure}[t]
\begin{centering}
\includegraphics[clip,width=0.98\columnwidth]{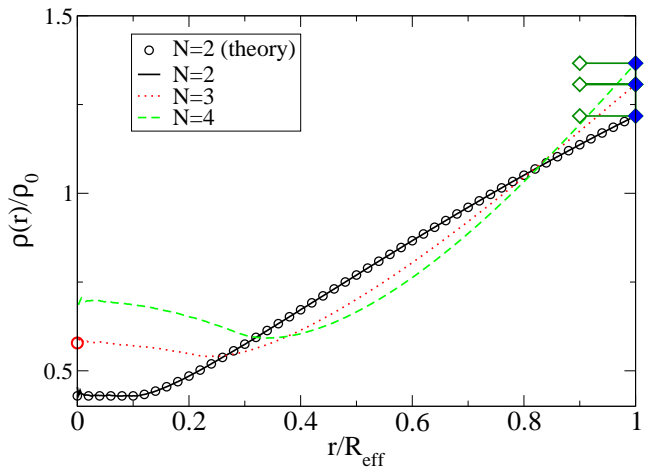}
\par\end{centering}

\protect\caption{(color online) Scaled density profiles for $\rho_{0}=0.35$ for $2$,
$3$ and $4$ particles. The exact analytical result for $N=2$ is
depicted with open symbols. A rich behavior with a main enhancement
of density close to the wall is observed. While for $N=2$ the bulk-like
plateau region at the center of the cavity is fairly constant, for
$3$ and $4$ particles presents a convexity, which leads to a minimum
at some intermediate point. The values of $\beta P_{w}/\rho_{0}$
extracted from both, theory and simulations, are presented in diamonds.
\label{fig:dens_prof_035_234}}
 
\end{figure}
In Figure \ref{fig:dens_prof_035_234}, we present details of the
scaled density profiles for an intermediate value of number density
($\rho_{0}=0.35\sigma^{-3}$) and $N=2,3,4$. Open circles for $N=2$
and for $N=3$ at $r=0$, correspond to exact theoretical results.
Diamonds show the scaled wall pressure $\beta P_{w}/\rho_{0}$, that
should be identical to the scaled density $\rho(r)/\rho_{0}$ at the
wall in accord with contact theorem {[}Eq. (\ref{eq:rhoscontact}){]}.
Closed diamonds show theoretical results. The case of $N=4$, was
calculated with the simulation-assisted approach. The figure also
shows, with open diamonds, the scaled wall pressure $\beta P_{w}/\rho_{0}$
obtained with MD. They were shifted to the left and a horizontal line
was draw to guide the eyes. The agreement with the contact theorem
and between theory and simulation are, again, both excellent. These
results are very important to validate the simulation-assisted fitting,
used for quasi-exact calculations in the case $N=4.$ The profiles
have similar features, but they have significant differences as changes
the number of particles. A main common feature is the enhancement
of density towards the wall. This can be understood as an effective
attraction produced by the translational entropy maximization. The
slightly enhanced presence of particles close to the wall, reduces
the mean excluded volume, which increases the number of possible configurations,
and therefore the entropy of the system. This can be also considered
as a special case of the depletion interaction\cite{Krauth_06}. Another
interesting feature, noticeable for $N=3$ and $4$ is that the density
profile presents a local minimum, and also structure in the central
region of the cavity. This features in the density profile were also
observed by White et. al. for Monte-Carlo and density functional theory
calculations of $N=6,8,10$\cite{White_2000}. 

\begin{figure}[t]
\begin{centering}
\includegraphics[clip,width=0.98\columnwidth]{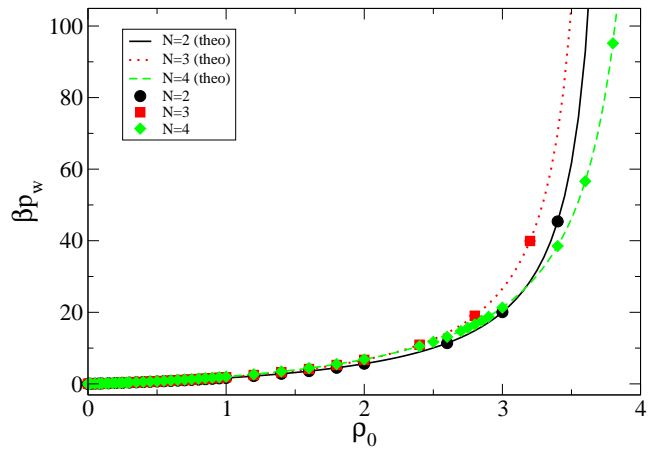}
\par\end{centering}

\protect\caption{(color online) Wall pressure versus number density. Lines for $N=2,3$
correspond to exact results, while for $N=4$ it corresponds to simulation-assisted
result. Symbols show simulation results.\label{fig:Wall_pressure}}
\end{figure}
In Figure \ref{fig:Wall_pressure}, we show pressure on the wall for
$N=$$2$,$3$,$4$ particles as a function of mean number density.
For $N=2,3$ the high degree of coincidence between simulation (circles
and squares) and theoretical results (continuous and dotted lines)
can be observed. In the case of four particles, one can observe that
the simulation results (diamonds) are well described by that found
using the simulation-assisted approach (dashed line).

\begin{figure}[t]
\begin{centering}
\includegraphics[clip,width=0.98\columnwidth]{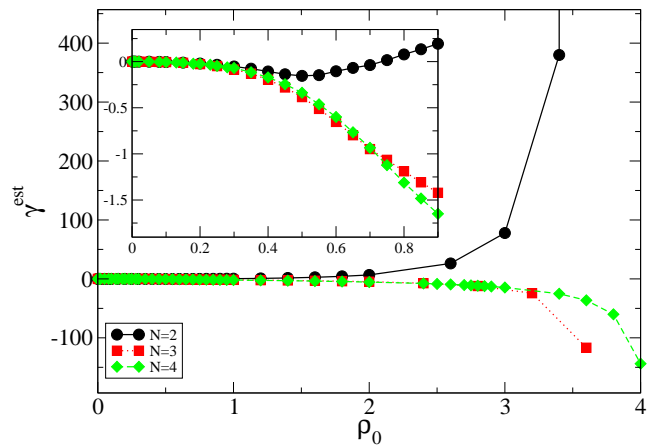}
\par\end{centering}

\protect\caption{(color online) Estimated system-substrate surface tension versus number
density for $N=2,3$ and $4$ at $T=2$. This quantity can be regarded
as a measure of the interface tension for these systems of few particles,
far from the thermodynamic limit. Symbols show simulation results
and lines between symbols are a guide for the eyes. For comparison,
the curve in full line (blue) shows the surface tension for the bulk
HS system in contact with a planar wall.\label{fig:surface_tension}}
\end{figure}
In Figure \ref{fig:surface_tension} we show an estimator of the system-substrate
surface tension $\gamma^{est}\equiv\left(P_{\textrm{center}}-P_{w}\right)R_{\textit{{\rm eff}}}/2$,
as a function of number density, based on the Laplace equation \cite{Henderson_1983}.
At low densities $\gamma^{est}$ is close to 0 for all the number
of particles considered, showing a vanishingly small energetic cost
to confine the system. The inset in Fig. \ref{fig:surface_tension}
is a zoom showing how the curves depart from 0 upon increasing confinement.
A local minimum of $\gamma^{est}$ is observed for $N=2$. At higher
number densities, there is an interesting important difference among
the cases $N=2$ and $N=3,4$. This qualitatively different behavior
is a consequence of the different packing features, when the system
becomes strongly constrained. There are limiting number densities
$\rho_{0}^{c}=1.10266$ ($N=3)$ and $\rho_{0}^{c}=1.47021$ ($N=4$)
such that $P_{\textrm{center}}$ vanishes for $\rho_{0}>\rho_{0}^{c}$.
In these cases, there are no possible configurations of the system
such that the segment joining the center of two colliding particles,
also passes through the center of the cavity. This geometrical constraint
becomes evident at lower densities $\rho_{0}\lesssim\rho_{0}^{c}$,
where there are few configurations that produce non-vanishing pressure
in the center. At higher densities, the pressure difference diverges
as $-P_{w}$. The case of two particles does not has this geometrical
restriction and then $P_{\textrm{center}}$ diverges with increasing
density. In this case, the opposite mechanism takes place: when the
caging is high and a lower total number of configurations is possible,
a progressively higher number of them have inter-particle segment
going through the center of the cavity. This divergence of $P_{\textrm{center}}$
is stronger than that of $P_{w}$ and thus $\gamma^{est}$ diverges
to $+\infty$ for $N=2$. For comparison, the system-substrate surface
tension of the infinitely extended HS system in contact with a planar
hard-wall is also shown\cite{Urrutia_2014}. It is negative and its
behavior follows that found for $N=2,3,4$ at very low densities.
The attained maximum density is in this case $\rho_{0}=1.4142$, the
maximum possible density of the bulk-solid system.

\begin{figure}[t]
\begin{centering}
\includegraphics[clip,width=0.98\columnwidth]{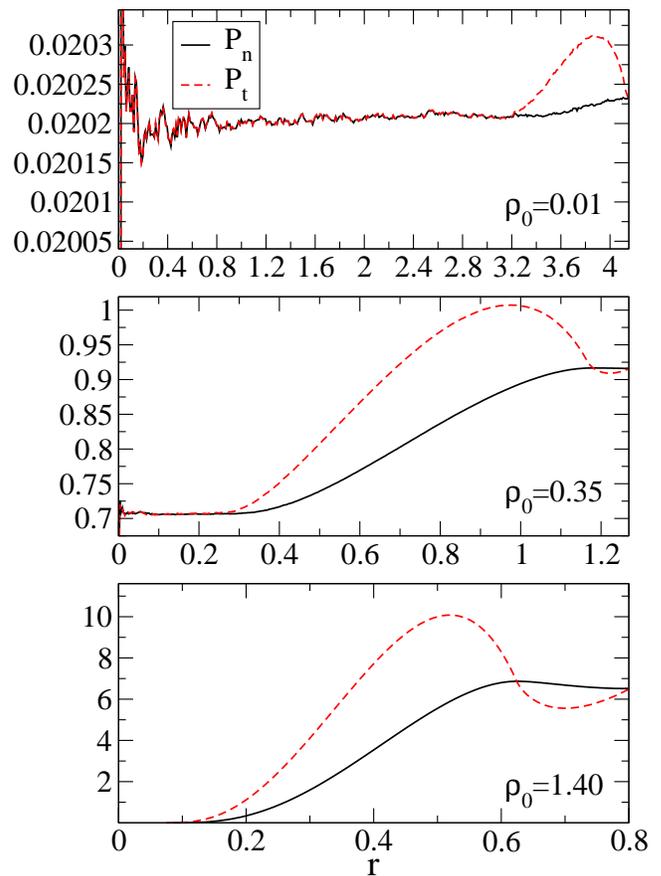}
\par\end{centering}

\protect\caption{(color online) Pressure tensor profile across the spherical cavity
for $N=3$ at $T=2$ for three distinctive number densities $\rho_{0}$.
Their components were chosen as $P_{n}$, normal to the wall and $P_{t}$,
tangential to the wall. The behavior is illustrated with three cases
of low (upper panel), medium (center panel) and high (lower panel)
number densities. Note the different scales, spanning four orders
of magnitude, of pressure values.\label{fig:Pressure-tensor-profile}}
\end{figure}
We also studied the pressure tensor profile across the spherical cavity.
As an example, Figure \ref{fig:Pressure-tensor-profile} presents
the profiles of the normal and tangential components of the pressure
tensor for $N=3$. In the center of the cavity $r\sim0$ for low ($\rho_{0}=0.01$)
and intermediate ($\rho_{0}=0.35$) densities, the behavior $P_{t}=P_{n}\neq0$
is representative of a zone of bulk-like behavior. In the interior
of an homogeneous and isotropic liquid all the components of the pressure
tensor have equal magnitude. Upon increasing number density, the bulk-liquid
region is reduced towards the center of the cavity and the region
in which there are important differences between the normal and tangential
components of the pressure tensor increases. This can be regarded
as the interfacial-like region of the cavity in which the differences
in the components of the pressure give rise to a non-vanishing contribution
to the surface tension as was shown in Fig. \ref{fig:surface_tension}.

At high densities the pressure in the central region is usually zero
($P_{t}=P_{n}=0$), as it is observed in the panel for $\rho_{0}=1.40$
in Fig. \ref{fig:Pressure-tensor-profile}. In line with the surface
tension, this can be rationalized by thinking in an event of colliding
particles. For a given pair of colliding particles, the contribution
to the local pressure is assigned along the straight line which connects
the pair. If, due to geometrical restrictions of confinement and excluded
volume, all of these lines lay out of the central region, the contribution
to the local pressure tensor components vanishes. At these high densities,
there is no bulk behavior at all, and the center of the particles
are not allowed to move through the center of the spherical cavity.
It can also be the case (not shown) in which the density in the central
region is 0 and the local pressure have a non-vanishing value. This
corresponds to a high-packing case, in which the center of particles
cannot be close to $r\sim0$, but the lines connecting a colliding
pair can pass through the central region. This happens, for example,
in the case of $N=2$ and $R_{\textrm{eff}}<1$ or number density
$\rho_{0}>\frac{6}{4\pi}\simeq0.48$.%

\begin{figure}[t]
\begin{centering}
\includegraphics[clip,width=0.98\columnwidth]{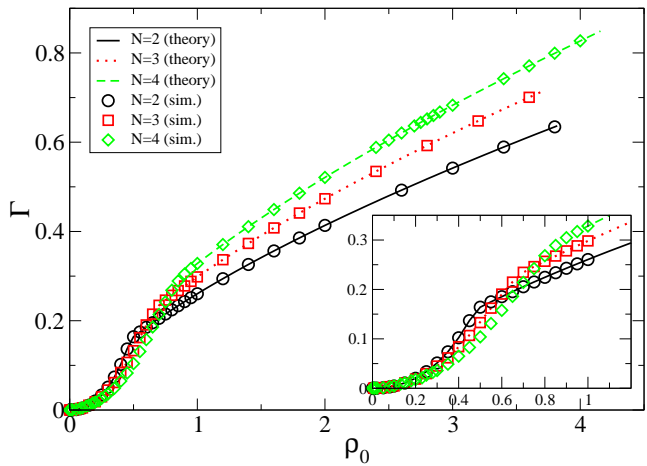}
\par\end{centering}

\protect\caption{(color online) Adsorption versus number density. Lines for $N=2,3$
correspond to exact results, while for $N=4$, the data corresponds
to simulation-assisted results. Symbols show the values of adsorption
obtained by simulations. The inset shows a zoom from 0 to intermediate
number densities. It is worth-mentioning that the adsorption for $N=1$
is zero for any density.\label{fig:Adsorption}}
\end{figure}
In Figure \ref{fig:Adsorption} the adsorption $\Gamma=\left(N-\rho_{\textrm{center}}V\right)/A$
is shown. The data from simulation is depicted with symbols and the
theoretical results with lines. This quantity measures the excess
of number density in the surface, as compared to a gas with mean number
density $\rho_{\textrm{center}}$. The curves present theoretical
results for $N=2,3$ and $4$. For $N=4$, adsorption is drawn from
the $\rho_{0}$ value such that that $\rho_{\textrm{center}}=0$.
An excellent agreement between theory and simulation is observed.
At low densities the adsorption tends to 0 for the three cases we
studied, in the limit of ideal gas. Then a fast rise of adsorption
is observed, in which the curves for different number of particles
are concave and cross each other (see inset of Fig. \ref{fig:Adsorption}).
The concave region ends up when $\rho_{\textrm{center}}$ vanishes.
At high densities, the adsorption is higher for greater number of
particles with a monotonic behavior in all the cases. This goes in
line with with the behavior of the surface tension for $N=3,4$ (see
Fig. \ref{fig:surface_tension}), in which a smaller surface tension
gives rise to a higher adsorption in the walls of the cavity.

\begin{figure}[t]
\centering{}\includegraphics[clip,width=0.98\columnwidth]{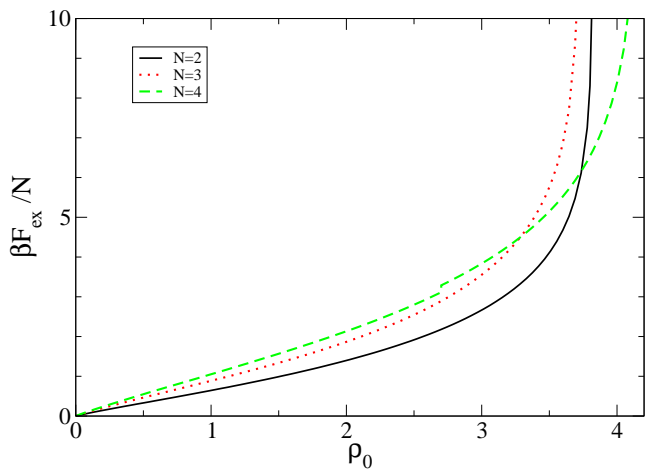}\protect\caption{(color online) Excess free energy per particle as function of number
density for different number of hard spheres. Continuous lines for
$N=2,3$ correspond to exact results while for $N=4$ it corresponds
to simulation-assisted result. The excess free energy is identically
zero for $N=1$.\label{fig:FreeEnergy}}
\end{figure}
Figure \ref{fig:FreeEnergy} presents the (Helmholtz) excess Free
energy $F_{\textrm{ex}}=F-F_{\textrm{ideal}}$ per particle for the
three cases considered in this work. The exact analytical results
are given for $N=2,3$ and the simultation-assisted result is provided
for $N=4$. It is worth noticing that this curve is very difficult
to obtain directly by simulations for such a wide range of number
densities. Within the simulation approach, this would mean performing
a numerical thermodynamic integration for each point of each curve,
with typically ten to twenty runs per free energy value\cite{frenkel-smit}.
The combined simulation-assisted approach provides this thermal quantity
in all the relevant range of number densities. The excess free energy
per particle for a hard sphere system is, of course, $F_{\textrm{ex}}=-TS_{\textrm{ex}}$.
Therefore, it gives an idea of how the entropy is reduced for the
different number of particles, when the number density increases in
the spherical cavity, and the available number of possible states
of the system reduces. The free energy curves cross each other, showing
the influence of the confinement in the free energy of a few particles
at high densities.

\section{Conclusions\label{sec:Conclusions}}

In this work we studied the properties of 2, 3 and 4 hard spheres
confined in a spherical cavity at constant temperature. We used analytical
calculations and event driven molecular dynamics simulations to study
the system for a wide range of number densities. We described the
system with variables and concepts widely used in thermodynamics and
macroscopic theory of liquids, and provided a comprehensive characterization,
covering very different physical regimes from ideal gas to highly-packed.
The results we obtained are interesting in the context of theory of
liquids, but also for new developments in material science and synthesis
of hollow nano-particles and rattles. In this minimal system, only
the excluded volume is present, and changes in free energy can only
be due to variations of translational entropy. This range of phenomena
is common to any experimental system and will have a role worth-understanding,
independently of the characteristics of the confining cavity or the
nature of the confined colloids or nano-particles. We characterized
the density and pressure tensor components profiles as a function
of number density and analyzed the distinctive behavior of an estimation
of the surface tension for different number of particles. The adsorption
was also characterized. In many quantities we found a rich and non-trivial
behavior, which was ultimately rationalized with the geometrical differences
of these few body systems. We compared carefully analytical and simulation
results whenever possible, to validate both methodologies. A simulation-assisted
approach was also used to obtain an analytical quasi-exact expression
for the free energy and pressure on the wall for the case $N=4$,
where a full analytical calculation was not found. Using this, we
calculated and compared the free energy of the system for all the
studied number of particles. 

This study will be extended by adding attraction among particles,
which is another important feature present in real systems. In this
context of minimal models, this will be achieved by using square-well
potentials, which in addition to the hard wall repulsion, have a short
range attraction. Also the properties of the wall can be modified
to model more realistic systems, for example adding a wall attraction.
This could be an interesting direction to understand these highly
confined systems.
\begin{acknowledgments}
We thank Iván Paganini for fruitful discussions. The financial support
through grants PICT-2011-1887, PIP 112-200801-00403, PIP-0546/10,
UBACyT 20020100200156, INN-CNEA 2011 and MINCYT-DAAD 11/09, are gratefully
acknowledged.
\end{acknowledgments}

\bibliographystyle{apsrev4-1}

\begin{thebibliography}{46}%
\makeatletter
\providecommand \@ifxundefined [1]{%
 \@ifx{#1\undefined}
}%
\providecommand \@ifnum [1]{%
 \ifnum #1\expandafter \@firstoftwo
 \else \expandafter \@secondoftwo
 \fi
}%
\providecommand \@ifx [1]{%
 \ifx #1\expandafter \@firstoftwo
 \else \expandafter \@secondoftwo
 \fi
}%
\providecommand \natexlab [1]{#1}%
\providecommand \enquote  [1]{``#1''}%
\providecommand \bibnamefont  [1]{#1}%
\providecommand \bibfnamefont [1]{#1}%
\providecommand \citenamefont [1]{#1}%
\providecommand \href@noop [0]{\@secondoftwo}%
\providecommand \href [0]{\begingroup \@sanitize@url \@href}%
\providecommand \@href[1]{\@@startlink{#1}\@@href}%
\providecommand \@@href[1]{\endgroup#1\@@endlink}%
\providecommand \@sanitize@url [0]{\catcode `\\12\catcode `\$12\catcode
  `\&12\catcode `\#12\catcode `\^12\catcode `\_12\catcode `\%12\relax}%
\providecommand \@@startlink[1]{}%
\providecommand \@@endlink[0]{}%
\providecommand \url  [0]{\begingroup\@sanitize@url \@url }%
\providecommand \@url [1]{\endgroup\@href {#1}{\urlprefix }}%
\providecommand \urlprefix  [0]{URL }%
\providecommand \Eprint [0]{\href }%
\providecommand \doibase [0]{http://dx.doi.org/}%
\providecommand \selectlanguage [0]{\@gobble}%
\providecommand \bibinfo  [0]{\@secondoftwo}%
\providecommand \bibfield  [0]{\@secondoftwo}%
\providecommand \translation [1]{[#1]}%
\providecommand \BibitemOpen [0]{}%
\providecommand \bibitemStop [0]{}%
\providecommand \bibitemNoStop [0]{.\EOS\space}%
\providecommand \EOS [0]{\spacefactor3000\relax}%
\providecommand \BibitemShut  [1]{\csname bibitem#1\endcsname}%
\let\auto@bib@innerbib\@empty
\bibitem [{\citenamefont {Tan}\ \emph {et~al.}(2010)\citenamefont {Tan},
  \citenamefont {Chen}, \citenamefont {Liu},\ and\ \citenamefont
  {Tang}}]{Tan_2010}%
  \BibitemOpen
  \bibfield  {author} {\bibinfo {author} {\bibfnamefont {L.}~\bibnamefont
  {Tan}}, \bibinfo {author} {\bibfnamefont {D.}~\bibnamefont {Chen}}, \bibinfo
  {author} {\bibfnamefont {H.}~\bibnamefont {Liu}}, \ and\ \bibinfo {author}
  {\bibfnamefont {F.}~\bibnamefont {Tang}},\ }\href {\doibase
  10.1002/adma.201002277} {\bibfield  {journal} {\bibinfo  {journal} {Advanced
  Materials}\ }\textbf {\bibinfo {volume} {22}},\ \bibinfo {pages} {4885}
  (\bibinfo {year} {2010})}\BibitemShut {NoStop}%
\bibitem [{\citenamefont {Tang}\ \emph {et~al.}(2012)\citenamefont {Tang},
  \citenamefont {Li},\ and\ \citenamefont {Chen}}]{Tang_12}%
  \BibitemOpen
  \bibfield  {author} {\bibinfo {author} {\bibfnamefont {F.}~\bibnamefont
  {Tang}}, \bibinfo {author} {\bibfnamefont {L.}~\bibnamefont {Li}}, \ and\
  \bibinfo {author} {\bibfnamefont {D.}~\bibnamefont {Chen}},\ }\href {\doibase
  10.1002/adma.201104763} {\bibfield  {journal} {\bibinfo  {journal} {Advanced
  Materials}\ }\textbf {\bibinfo {volume} {24}},\ \bibinfo {pages} {1504}
  (\bibinfo {year} {2012})}\BibitemShut {NoStop}%
\bibitem [{\citenamefont {Chen}\ \emph {et~al.}(2010)\citenamefont {Chen},
  \citenamefont {Chen}, \citenamefont {Guo}, \citenamefont {He}, \citenamefont
  {Chen}, \citenamefont {Zhou}, \citenamefont {Feng},\ and\ \citenamefont
  {Shi}}]{Chen_10}%
  \BibitemOpen
  \bibfield  {author} {\bibinfo {author} {\bibfnamefont {Y.}~\bibnamefont
  {Chen}}, \bibinfo {author} {\bibfnamefont {H.}~\bibnamefont {Chen}}, \bibinfo
  {author} {\bibfnamefont {L.}~\bibnamefont {Guo}}, \bibinfo {author}
  {\bibfnamefont {Q.}~\bibnamefont {He}}, \bibinfo {author} {\bibfnamefont
  {F.}~\bibnamefont {Chen}}, \bibinfo {author} {\bibfnamefont {J.}~\bibnamefont
  {Zhou}}, \bibinfo {author} {\bibfnamefont {J.}~\bibnamefont {Feng}}, \ and\
  \bibinfo {author} {\bibfnamefont {J.}~\bibnamefont {Shi}},\ }\href {\doibase
  10.1021/nn901398j} {\bibfield  {journal} {\bibinfo  {journal} {ACS Nano}\
  }\textbf {\bibinfo {volume} {4}},\ \bibinfo {pages} {529} (\bibinfo {year}
  {2010})}\BibitemShut {NoStop}%
\bibitem [{\citenamefont {Yang}\ \emph {et~al.}(2008)\citenamefont {Yang},
  \citenamefont {Lee}, \citenamefont {Kang}, \citenamefont {Lee}, \citenamefont
  {Suh}, \citenamefont {Yoon}, \citenamefont {Huh},\ and\ \citenamefont
  {Haam}}]{Yang_08}%
  \BibitemOpen
  \bibfield  {author} {\bibinfo {author} {\bibfnamefont {J.}~\bibnamefont
  {Yang}}, \bibinfo {author} {\bibfnamefont {J.}~\bibnamefont {Lee}}, \bibinfo
  {author} {\bibfnamefont {J.}~\bibnamefont {Kang}}, \bibinfo {author}
  {\bibfnamefont {K.}~\bibnamefont {Lee}}, \bibinfo {author} {\bibfnamefont
  {J.-S.}\ \bibnamefont {Suh}}, \bibinfo {author} {\bibfnamefont {H.-G.}\
  \bibnamefont {Yoon}}, \bibinfo {author} {\bibfnamefont {Y.-M.}\ \bibnamefont
  {Huh}}, \ and\ \bibinfo {author} {\bibfnamefont {S.}~\bibnamefont {Haam}},\
  }\href {\doibase 10.1021/la701688t} {\bibfield  {journal} {\bibinfo
  {journal} {Langmuir}\ }\textbf {\bibinfo {volume} {24}},\ \bibinfo {pages}
  {3417} (\bibinfo {year} {2008})}\BibitemShut {NoStop}%
\bibitem [{\citenamefont {Wang}\ \emph {et~al.}(2011)\citenamefont {Wang},
  \citenamefont {Chai}, \citenamefont {Wang}, \citenamefont {Li}, \citenamefont
  {Liu}, \citenamefont {Zhang}, \citenamefont {Su},\ and\ \citenamefont
  {Liao}}]{Wang_11}%
  \BibitemOpen
  \bibfield  {author} {\bibinfo {author} {\bibfnamefont {T.-T.}\ \bibnamefont
  {Wang}}, \bibinfo {author} {\bibfnamefont {F.}~\bibnamefont {Chai}}, \bibinfo
  {author} {\bibfnamefont {C.-G.}\ \bibnamefont {Wang}}, \bibinfo {author}
  {\bibfnamefont {L.}~\bibnamefont {Li}}, \bibinfo {author} {\bibfnamefont
  {H.-Y.}\ \bibnamefont {Liu}}, \bibinfo {author} {\bibfnamefont {L.-Y.}\
  \bibnamefont {Zhang}}, \bibinfo {author} {\bibfnamefont {Z.-M.}\ \bibnamefont
  {Su}}, \ and\ \bibinfo {author} {\bibfnamefont {Y.}~\bibnamefont {Liao}},\
  }\href {\doibase http://dx.doi.org/10.1016/j.jcis.2011.02.023} {\bibfield
  {journal} {\bibinfo  {journal} {Journal of Colloid and Interface Science}\
  }\textbf {\bibinfo {volume} {358}},\ \bibinfo {pages} {109 } (\bibinfo {year}
  {2011})}\BibitemShut {NoStop}%
\bibitem [{\citenamefont {Liu}\ \emph {et~al.}(2010)\citenamefont {Liu},
  \citenamefont {Qiao}, \citenamefont {Budi~Hartono},\ and\ \citenamefont
  {Lu}}]{Liu_10}%
  \BibitemOpen
  \bibfield  {author} {\bibinfo {author} {\bibfnamefont {J.}~\bibnamefont
  {Liu}}, \bibinfo {author} {\bibfnamefont {S.~Z.}\ \bibnamefont {Qiao}},
  \bibinfo {author} {\bibfnamefont {S.}~\bibnamefont {Budi~Hartono}}, \ and\
  \bibinfo {author} {\bibfnamefont {G.~K.}\ \bibnamefont {Lu}},\ }\href
  {\doibase 10.1002/ange.201001252} {\bibfield  {journal} {\bibinfo  {journal}
  {Angewandte Chemie}\ }\textbf {\bibinfo {volume} {122}},\ \bibinfo {pages}
  {5101} (\bibinfo {year} {2010})}\BibitemShut {NoStop}%
\bibitem [{\citenamefont {Dinsmore}\ \emph {et~al.}(1998)\citenamefont
  {Dinsmore}, \citenamefont {Wong}, \citenamefont {Nelson},\ and\ \citenamefont
  {Yodh}}]{Dinsmore_1998}%
  \BibitemOpen
  \bibfield  {author} {\bibinfo {author} {\bibfnamefont {A.~D.}\ \bibnamefont
  {Dinsmore}}, \bibinfo {author} {\bibfnamefont {D.~T.}\ \bibnamefont {Wong}},
  \bibinfo {author} {\bibfnamefont {P.}~\bibnamefont {Nelson}}, \ and\ \bibinfo
  {author} {\bibfnamefont {A.~G.}\ \bibnamefont {Yodh}},\ }\href {\doibase
  10.1103/PhysRevLett.80.409} {\bibfield  {journal} {\bibinfo  {journal} {Phys.
  Rev. Lett.}\ }\textbf {\bibinfo {volume} {80}},\ \bibinfo {pages} {409}
  (\bibinfo {year} {1998})}\BibitemShut {NoStop}%
\bibitem [{\citenamefont {Statt}\ \emph {et~al.}(2012)\citenamefont {Statt},
  \citenamefont {Winkler}, \citenamefont {Virnau},\ and\ \citenamefont
  {Binder}}]{Statt_2012}%
  \BibitemOpen
  \bibfield  {author} {\bibinfo {author} {\bibfnamefont {A.}~\bibnamefont
  {Statt}}, \bibinfo {author} {\bibfnamefont {A.}~\bibnamefont {Winkler}},
  \bibinfo {author} {\bibfnamefont {P.}~\bibnamefont {Virnau}}, \ and\ \bibinfo
  {author} {\bibfnamefont {K.}~\bibnamefont {Binder}},\ }\href
  {http://stacks.iop.org/0953-8984/24/i=46/a=464122} {\bibfield  {journal}
  {\bibinfo  {journal} {Journal of Physics: Condensed Matter}\ }\textbf
  {\bibinfo {volume} {24}},\ \bibinfo {pages} {464122} (\bibinfo {year}
  {2012})}\BibitemShut {NoStop}%
\bibitem [{\citenamefont {Huang}\ \emph {et~al.}(2013)\citenamefont {Huang},
  \citenamefont {Yoon},\ and\ \citenamefont {Kwak}}]{Huang_2013}%
  \BibitemOpen
  \bibfield  {author} {\bibinfo {author} {\bibfnamefont {H.~C.}\ \bibnamefont
  {Huang}}, \bibinfo {author} {\bibfnamefont {Y.~J.}\ \bibnamefont {Yoon}}, \
  and\ \bibinfo {author} {\bibfnamefont {S.~K.}\ \bibnamefont {Kwak}},\ }\href
  {\doibase 10.1080/00268976.2013.781694} {\bibfield  {journal} {\bibinfo
  {journal} {Molecular Physics}\ }\textbf {\bibinfo {volume} {111}},\ \bibinfo
  {pages} {3283} (\bibinfo {year} {2013})}\BibitemShut {NoStop}%
\bibitem [{\citenamefont {Roth}(2010)}]{Roth_2010}%
  \BibitemOpen
  \bibfield  {author} {\bibinfo {author} {\bibfnamefont {R.}~\bibnamefont
  {Roth}},\ }\href {\doibase 10.1088/0953-8984/22/6/063102} {\bibfield
  {journal} {\bibinfo  {journal} {Journal of Physics: Condensed Matter}\
  }\textbf {\bibinfo {volume} {22}},\ \bibinfo {pages} {063102} (\bibinfo
  {year} {2010})}\BibitemShut {NoStop}%
\bibitem [{\citenamefont {Zykova-Timan}\ \emph {et~al.}(2010)\citenamefont
  {Zykova-Timan}, \citenamefont {Horbach},\ and\ \citenamefont
  {Binder}}]{Zykova_2010}%
  \BibitemOpen
  \bibfield  {author} {\bibinfo {author} {\bibfnamefont {T.}~\bibnamefont
  {Zykova-Timan}}, \bibinfo {author} {\bibfnamefont {J.}~\bibnamefont
  {Horbach}}, \ and\ \bibinfo {author} {\bibfnamefont {K.}~\bibnamefont
  {Binder}},\ }\href {\doibase 10.1063/1.3455504} {\bibfield  {journal}
  {\bibinfo  {journal} {The Journal of Chemical Physics}\ }\textbf {\bibinfo
  {volume} {133}},\ \bibinfo {eid} {014705} (\bibinfo {year}
  {2010})}\BibitemShut {NoStop}%
\bibitem [{\citenamefont {Winkler}\ \emph {et~al.}(2013)\citenamefont
  {Winkler}, \citenamefont {Statt}, \citenamefont {Virnau},\ and\ \citenamefont
  {Binder}}]{Winkler_2013}%
  \BibitemOpen
  \bibfield  {author} {\bibinfo {author} {\bibfnamefont {A.}~\bibnamefont
  {Winkler}}, \bibinfo {author} {\bibfnamefont {A.}~\bibnamefont {Statt}},
  \bibinfo {author} {\bibfnamefont {P.}~\bibnamefont {Virnau}}, \ and\ \bibinfo
  {author} {\bibfnamefont {K.}~\bibnamefont {Binder}},\ }\href {\doibase
  10.1103/PhysRevE.87.032307} {\bibfield  {journal} {\bibinfo  {journal} {Phys.
  Rev. E}\ }\textbf {\bibinfo {volume} {87}},\ \bibinfo {pages} {032307}
  (\bibinfo {year} {2013})}\BibitemShut {NoStop}%
\bibitem [{\citenamefont {Alder}\ and\ \citenamefont
  {Wainwright}(1957)}]{Alder_1957}%
  \BibitemOpen
  \bibfield  {author} {\bibinfo {author} {\bibfnamefont {B.~J.}\ \bibnamefont
  {Alder}}\ and\ \bibinfo {author} {\bibfnamefont {T.~E.}\ \bibnamefont
  {Wainwright}},\ }\href {\doibase 10.1063/1.1743957} {\bibfield  {journal}
  {\bibinfo  {journal} {The Journal of Chemical Physics}\ }\textbf {\bibinfo
  {volume} {27}},\ \bibinfo {pages} {1208} (\bibinfo {year}
  {1957})}\BibitemShut {NoStop}%
\bibitem [{\citenamefont {Alder}\ \emph {et~al.}(1963)\citenamefont {Alder},
  \citenamefont {Hoover},\ and\ \citenamefont {Wainwright}}]{Alder_1963}%
  \BibitemOpen
  \bibfield  {author} {\bibinfo {author} {\bibfnamefont {B.~J.}\ \bibnamefont
  {Alder}}, \bibinfo {author} {\bibfnamefont {W.~G.}\ \bibnamefont {Hoover}}, \
  and\ \bibinfo {author} {\bibfnamefont {T.~E.}\ \bibnamefont {Wainwright}},\
  }\href {\doibase 10.1103/PhysRevLett.11.241} {\bibfield  {journal} {\bibinfo
  {journal} {Phys. Rev. Lett.}\ }\textbf {\bibinfo {volume} {11}},\ \bibinfo
  {pages} {241} (\bibinfo {year} {1963})}\BibitemShut {NoStop}%
\bibitem [{\citenamefont {Alder}\ \emph {et~al.}(1968)\citenamefont {Alder},
  \citenamefont {Hoover},\ and\ \citenamefont {Young}}]{Alder_1968}%
  \BibitemOpen
  \bibfield  {author} {\bibinfo {author} {\bibfnamefont {B.~J.}\ \bibnamefont
  {Alder}}, \bibinfo {author} {\bibfnamefont {W.~G.}\ \bibnamefont {Hoover}}, \
  and\ \bibinfo {author} {\bibfnamefont {D.~A.}\ \bibnamefont {Young}},\ }\href
  {\doibase 10.1063/1.1670653} {\bibfield  {journal} {\bibinfo  {journal} {The
  Journal of Chemical Physics}\ }\textbf {\bibinfo {volume} {49}},\ \bibinfo
  {pages} {3688} (\bibinfo {year} {1968})}\BibitemShut {NoStop}%
\bibitem [{\citenamefont {Macpherson}\ \emph {et~al.}(1987)\citenamefont
  {Macpherson}, \citenamefont {Carignan},\ and\ \citenamefont
  {Vladimiroff}}]{Macpherson_1987}%
  \BibitemOpen
  \bibfield  {author} {\bibinfo {author} {\bibfnamefont {A.~K.}\ \bibnamefont
  {Macpherson}}, \bibinfo {author} {\bibfnamefont {Y.~P.}\ \bibnamefont
  {Carignan}}, \ and\ \bibinfo {author} {\bibfnamefont {T.}~\bibnamefont
  {Vladimiroff}},\ }\href {\doibase 10.1063/1.453189} {\bibfield  {journal}
  {\bibinfo  {journal} {Journal of Chemical Physics}\ }\textbf {\bibinfo
  {volume} {87}},\ \bibinfo {pages} {1768} (\bibinfo {year}
  {1987})}\BibitemShut {NoStop}%
\bibitem [{\citenamefont {Gonz\'{a}lez}\ \emph {et~al.}(1997)\citenamefont
  {Gonz\'{a}lez}, \citenamefont {White}, \citenamefont {Rom\'{a}n},
  \citenamefont {Velasco},\ and\ \citenamefont {Evans}}]{Gonzalez_1997}%
  \BibitemOpen
  \bibfield  {author} {\bibinfo {author} {\bibfnamefont {A.}~\bibnamefont
  {Gonz\'{a}lez}}, \bibinfo {author} {\bibfnamefont {J.~A.}\ \bibnamefont
  {White}}, \bibinfo {author} {\bibfnamefont {F.~L.}\ \bibnamefont
  {Rom\'{a}n}}, \bibinfo {author} {\bibfnamefont {S.}~\bibnamefont {Velasco}},
  \ and\ \bibinfo {author} {\bibfnamefont {R.}~\bibnamefont {Evans}},\ }\href
  {\doibase 10.1103/PhysRevLett.79.2466} {\bibfield  {journal} {\bibinfo
  {journal} {Phys. Rev. Lett.}\ }\textbf {\bibinfo {volume} {79}},\ \bibinfo
  {pages} {2466} (\bibinfo {year} {1997})}\BibitemShut {NoStop}%
\bibitem [{\citenamefont {Gonz\'{a}lez}\ \emph {et~al.}(2006)\citenamefont
  {Gonz\'{a}lez}, \citenamefont {White}, \citenamefont {Rom\'{a}n},\ and\
  \citenamefont {Velasco}}]{Gonzalez_2006}%
  \BibitemOpen
  \bibfield  {author} {\bibinfo {author} {\bibfnamefont {A.}~\bibnamefont
  {Gonz\'{a}lez}}, \bibinfo {author} {\bibfnamefont {J.~A.}\ \bibnamefont
  {White}}, \bibinfo {author} {\bibfnamefont {F.~L.}\ \bibnamefont
  {Rom\'{a}n}}, \ and\ \bibinfo {author} {\bibfnamefont {S.}~\bibnamefont
  {Velasco}},\ }\href {\doibase 10.1063/1.2227389} {\bibfield  {journal}
  {\bibinfo  {journal} {The Journal of Chemical Physics}\ }\textbf {\bibinfo
  {volume} {125}},\ \bibinfo {eid} {064703} (\bibinfo {year}
  {2006})}\BibitemShut {NoStop}%
\bibitem [{\citenamefont {Urrutia}(2008)}]{Urrutia_2008}%
  \BibitemOpen
  \bibfield  {author} {\bibinfo {author} {\bibfnamefont {I.}~\bibnamefont
  {Urrutia}},\ }\href {\doibase 10.1007/s10955-008-9513-3} {\bibfield
  {journal} {\bibinfo  {journal} {Journal of Statistical Physics}\ }\textbf
  {\bibinfo {volume} {131}},\ \bibinfo {pages} {597} (\bibinfo {year}
  {2008})}\BibitemShut {NoStop}%
\bibitem [{\citenamefont {Urrutia}\ and\ \citenamefont
  {Szybisz}(2010)}]{Urrutia_2010}%
  \BibitemOpen
  \bibfield  {author} {\bibinfo {author} {\bibfnamefont {I.}~\bibnamefont
  {Urrutia}}\ and\ \bibinfo {author} {\bibfnamefont {L.}~\bibnamefont
  {Szybisz}},\ }\href {\doibase 10.1063/1.3319560} {\bibfield  {journal}
  {\bibinfo  {journal} {Journal of Mathematical Physics}\ }\textbf {\bibinfo
  {volume} {51}},\ \bibinfo {pages} {033303} (\bibinfo {year} {2010})},\
  \bibinfo {note} {arXiv:0909.0246}\BibitemShut {NoStop}%
\bibitem [{\citenamefont {Urrutia}(2010)}]{Urrutia_2010b}%
  \BibitemOpen
  \bibfield  {author} {\bibinfo {author} {\bibfnamefont {I.}~\bibnamefont
  {Urrutia}},\ }\href {\doibase 10.1063/1.3469773} {\bibfield  {journal}
  {\bibinfo  {journal} {The Journal of Chemical Physics}\ }\textbf {\bibinfo
  {volume} {133}},\ \bibinfo {eid} {104503} (\bibinfo {year}
  {2010})}\BibitemShut {NoStop}%
\bibitem [{\citenamefont {Urrutia}(2011)}]{Urrutia_2011_b}%
  \BibitemOpen
  \bibfield  {author} {\bibinfo {author} {\bibfnamefont {I.}~\bibnamefont
  {Urrutia}},\ }\href {\doibase 10.1063/1.3609796} {\bibfield  {journal}
  {\bibinfo  {journal} {The Journal of Chemical Physics}\ }\textbf {\bibinfo
  {volume} {135}},\ \bibinfo {pages} {024511} (\bibinfo {year} {2011})},\
  \bibinfo {note} {erratum: ibid. 135(9), 099903 (2011)}\BibitemShut {NoStop}%
\bibitem [{\citenamefont {Blokhuis}(2013)}]{Blokhuis_2013}%
  \BibitemOpen
  \bibfield  {author} {\bibinfo {author} {\bibfnamefont {E.~M.}\ \bibnamefont
  {Blokhuis}},\ }\href {\doibase 10.1103/PhysRevE.87.022401} {\bibfield
  {journal} {\bibinfo  {journal} {Phys. Rev. E}\ }\textbf {\bibinfo {volume}
  {87}},\ \bibinfo {pages} {022401} (\bibinfo {year} {2013})}\BibitemShut
  {NoStop}%
\bibitem [{\citenamefont {Urrutia}(2014)}]{Urrutia_2014}%
  \BibitemOpen
  \bibfield  {author} {\bibinfo {author} {\bibfnamefont {I.}~\bibnamefont
  {Urrutia}},\ }\href {\doibase 10.1103/PhysRevE.89.032122} {\bibfield
  {journal} {\bibinfo  {journal} {Phys. Rev. E}\ }\textbf {\bibinfo {volume}
  {89}},\ \bibinfo {pages} {032122} (\bibinfo {year} {2014})}\BibitemShut
  {NoStop}%
\bibitem [{\citenamefont {Allen}\ and\ \citenamefont
  {Tildesley}(1987)}]{Allen_and_Tildesley}%
  \BibitemOpen
  \bibfield  {author} {\bibinfo {author} {\bibfnamefont {M.~P.}\ \bibnamefont
  {Allen}}\ and\ \bibinfo {author} {\bibfnamefont {D.~J.}\ \bibnamefont
  {Tildesley}},\ }\href@noop {} {\emph {\bibinfo {title} {Computer Simulations
  of Liquids}}}\ (\bibinfo  {publisher} {Clarendom Press, Oxford},\ \bibinfo
  {year} {1987})\BibitemShut {NoStop}%
\bibitem [{\citenamefont {Humphrey}\ \emph {et~al.}(1996)\citenamefont
  {Humphrey}, \citenamefont {Dalke},\ and\ \citenamefont {Schulten}}]{VMD}%
  \BibitemOpen
  \bibfield  {author} {\bibinfo {author} {\bibfnamefont {W.}~\bibnamefont
  {Humphrey}}, \bibinfo {author} {\bibfnamefont {A.}~\bibnamefont {Dalke}}, \
  and\ \bibinfo {author} {\bibfnamefont {K.}~\bibnamefont {Schulten}},\
  }\href@noop {} {\bibfield  {journal} {\bibinfo  {journal} {J. Mol. Graphics}\
  }\textbf {\bibinfo {volume} {14}},\ \bibinfo {pages} {33} (\bibinfo {year}
  {1996})}\BibitemShut {NoStop}%
\bibitem [{\citenamefont {Hill}(1956)}]{Hill1956}%
  \BibitemOpen
  \bibfield  {author} {\bibinfo {author} {\bibfnamefont {T.~L.}\ \bibnamefont
  {Hill}},\ }\href@noop {} {\emph {\bibinfo {title} {Statistical Mechanics}}}\
  (\bibinfo  {publisher} {Dover},\ \bibinfo {address} {New York},\ \bibinfo
  {year} {1956})\BibitemShut {NoStop}%
\bibitem [{\citenamefont {Urrutia}\ and\ \citenamefont
  {Castelletti}(2012)}]{Urrutia_2012}%
  \BibitemOpen
  \bibfield  {author} {\bibinfo {author} {\bibfnamefont {I.}~\bibnamefont
  {Urrutia}}\ and\ \bibinfo {author} {\bibfnamefont {G.}~\bibnamefont
  {Castelletti}},\ }\href {\doibase 10.1063/1.4729249} {\bibfield  {journal}
  {\bibinfo  {journal} {The Journal of Chemical Physics}\ }\textbf {\bibinfo
  {volume} {136}},\ \bibinfo {eid} {224509} (\bibinfo {year}
  {2012})}\BibitemShut {NoStop}%
\bibitem [{\citenamefont {Yang}\ \emph {et~al.}(2013)\citenamefont {Yang},
  \citenamefont {Schultz}, \citenamefont {Errington},\ and\ \citenamefont
  {Kofke}}]{Yang_2013}%
  \BibitemOpen
  \bibfield  {author} {\bibinfo {author} {\bibfnamefont {J.~H.}\ \bibnamefont
  {Yang}}, \bibinfo {author} {\bibfnamefont {A.~J.}\ \bibnamefont {Schultz}},
  \bibinfo {author} {\bibfnamefont {J.~R.}\ \bibnamefont {Errington}}, \ and\
  \bibinfo {author} {\bibfnamefont {D.~A.}\ \bibnamefont {Kofke}},\ }\href
  {\doibase 10.1063/1.4798456} {\bibfield  {journal} {\bibinfo  {journal} {The
  Journal of Chemical Physics}\ }\textbf {\bibinfo {volume} {138}},\ \bibinfo
  {eid} {134706} (\bibinfo {year} {2013})}\BibitemShut {NoStop}%
\bibitem [{\citenamefont {Hansen}\ and\ \citenamefont
  {McDonald}(2006)}]{Hansen2006}%
  \BibitemOpen
  \bibfield  {author} {\bibinfo {author} {\bibfnamefont {J.-P.}\ \bibnamefont
  {Hansen}}\ and\ \bibinfo {author} {\bibfnamefont {I.~R.}\ \bibnamefont
  {McDonald}},\ }\href@noop {} {\emph {\bibinfo {title} {Theory of simple
  liquids, 3rd Edition}}}\ (\bibinfo  {publisher} {Academic Press},\ \bibinfo
  {address} {Amsterdam},\ \bibinfo {year} {2006})\BibitemShut {NoStop}%
\bibitem [{\citenamefont {Blokhuis}\ and\ \citenamefont
  {Kuipers}(2007)}]{Blokhuis_2007}%
  \BibitemOpen
  \bibfield  {author} {\bibinfo {author} {\bibfnamefont {E.~M.}\ \bibnamefont
  {Blokhuis}}\ and\ \bibinfo {author} {\bibfnamefont {J.}~\bibnamefont
  {Kuipers}},\ }\href {\doibase 10.1063/1.2434161} {\bibfield  {journal}
  {\bibinfo  {journal} {The Journal of Chemical Physics}\ }\textbf {\bibinfo
  {volume} {126}},\ \bibinfo {eid} {054702} (\bibinfo {year}
  {2007})}\BibitemShut {NoStop}%
\bibitem [{\citenamefont {Munakata}\ and\ \citenamefont
  {Hu}(2002)}]{Munakata_2002}%
  \BibitemOpen
  \bibfield  {author} {\bibinfo {author} {\bibfnamefont {T.}~\bibnamefont
  {Munakata}}\ and\ \bibinfo {author} {\bibfnamefont {G.}~\bibnamefont {Hu}},\
  }\href {\doibase 10.1103/PhysRevE.65.066104} {\bibfield  {journal} {\bibinfo
  {journal} {Phys. Rev. E}\ }\textbf {\bibinfo {volume} {65}},\ \bibinfo
  {pages} {066104} (\bibinfo {year} {2002})}\BibitemShut {NoStop}%
\bibitem [{Note1()}]{Note1}%
  \BibitemOpen
  \bibinfo {note} {See for example p. 134 in the book by Hill \cite {Hill1956}
  or p. 28 in the book of Hansen and McDonald\cite {Hansen2006}}\BibitemShut
  {NoStop}%
\bibitem [{\citenamefont {Gonz\'{a}lez}\ \emph {et~al.}(1998)\citenamefont
  {Gonz\'{a}lez}, \citenamefont {White}, \citenamefont {Rom\'{a}n},\ and\
  \citenamefont {Evans}}]{Gonzalez_1998}%
  \BibitemOpen
  \bibfield  {author} {\bibinfo {author} {\bibfnamefont {A.}~\bibnamefont
  {Gonz\'{a}lez}}, \bibinfo {author} {\bibfnamefont {J.~A.}\ \bibnamefont
  {White}}, \bibinfo {author} {\bibfnamefont {F.~L.}\ \bibnamefont
  {Rom\'{a}n}}, \ and\ \bibinfo {author} {\bibfnamefont {R.}~\bibnamefont
  {Evans}},\ }\href {\doibase 10.1063/1.476961} {\bibfield  {journal} {\bibinfo
   {journal} {The Journal of Chemical Physics}\ }\textbf {\bibinfo {volume}
  {109}},\ \bibinfo {pages} {3637} (\bibinfo {year} {1998})}\BibitemShut
  {NoStop}%
\bibitem [{\citenamefont {H{\"u}nenberger}(2005)}]{Hunenberger_05}%
  \BibitemOpen
  \bibfield  {author} {\bibinfo {author} {\bibfnamefont {P.}~\bibnamefont
  {H{\"u}nenberger}},\ }\href@noop {} {\bibfield  {journal} {\bibinfo
  {journal} {Adv. Polym. Sci.}\ }\textbf {\bibinfo {volume} {173}},\ \bibinfo
  {pages} {105} (\bibinfo {year} {2005})}\BibitemShut {NoStop}%
\bibitem [{\citenamefont {Yong}\ and\ \citenamefont {Zhang}(2013)}]{Yong_2013}%
  \BibitemOpen
  \bibfield  {author} {\bibinfo {author} {\bibfnamefont {X.}~\bibnamefont
  {Yong}}\ and\ \bibinfo {author} {\bibfnamefont {L.~T.}\ \bibnamefont
  {Zhang}},\ }\href {\doibase 10.1063/1.4792202} {\bibfield  {journal}
  {\bibinfo  {journal} {The Journal of Chemical Physics}\ }\textbf {\bibinfo
  {volume} {138}},\ \bibinfo {eid} {084503} (\bibinfo {year}
  {2013})}\BibitemShut {NoStop}%
\bibitem [{\citenamefont {Tehver}\ \emph {et~al.}(1998)\citenamefont {Tehver},
  \citenamefont {Toigo}, \citenamefont {Koplik},\ and\ \citenamefont
  {Banavar}}]{Tehver_98}%
  \BibitemOpen
  \bibfield  {author} {\bibinfo {author} {\bibfnamefont {R.}~\bibnamefont
  {Tehver}}, \bibinfo {author} {\bibfnamefont {F.}~\bibnamefont {Toigo}},
  \bibinfo {author} {\bibfnamefont {J.}~\bibnamefont {Koplik}}, \ and\ \bibinfo
  {author} {\bibfnamefont {J.~R.}\ \bibnamefont {Banavar}},\ }\href {\doibase
  10.1103/PhysRevE.57.R17} {\bibfield  {journal} {\bibinfo  {journal} {Phys.
  Rev. E}\ }\textbf {\bibinfo {volume} {57}},\ \bibinfo {pages} {17} (\bibinfo
  {year} {1998})}\BibitemShut {NoStop}%
\bibitem [{\citenamefont {Inoue}\ \emph {et~al.}(2002)\citenamefont {Inoue},
  \citenamefont {Chen},\ and\ \citenamefont {Ohashi}}]{Inoue_02}%
  \BibitemOpen
  \bibfield  {author} {\bibinfo {author} {\bibfnamefont {Y.}~\bibnamefont
  {Inoue}}, \bibinfo {author} {\bibfnamefont {Y.}~\bibnamefont {Chen}}, \ and\
  \bibinfo {author} {\bibfnamefont {H.}~\bibnamefont {Ohashi}},\ }\href
  {\doibase 10.1023/A:1014550318814} {\bibfield  {journal} {\bibinfo  {journal}
  {Journal of Statistical Physics}\ }\textbf {\bibinfo {volume} {107}},\
  \bibinfo {pages} {85} (\bibinfo {year} {2002})}\BibitemShut {NoStop}%
\bibitem [{\citenamefont {Irving}\ and\ \citenamefont
  {Kirkwood}(1950)}]{Irving_1950}%
  \BibitemOpen
  \bibfield  {author} {\bibinfo {author} {\bibfnamefont {J.~H.}\ \bibnamefont
  {Irving}}\ and\ \bibinfo {author} {\bibfnamefont {J.~G.}\ \bibnamefont
  {Kirkwood}},\ }\href {\doibase 10.1063/1.1747782} {\bibfield  {journal}
  {\bibinfo  {journal} {The Journal of Chemical Physics}\ }\textbf {\bibinfo
  {volume} {18}},\ \bibinfo {pages} {817} (\bibinfo {year} {1950})}\BibitemShut
  {NoStop}%
\bibitem [{\citenamefont {Schofield}\ and\ \citenamefont
  {Henderson}(1982)}]{Schofield_1982}%
  \BibitemOpen
  \bibfield  {author} {\bibinfo {author} {\bibfnamefont {P.}~\bibnamefont
  {Schofield}}\ and\ \bibinfo {author} {\bibfnamefont {J.~R.}\ \bibnamefont
  {Henderson}},\ }\href {\doibase 10.1098/rspa.1982.0015} {\bibfield  {journal}
  {\bibinfo  {journal} {Proceedings of the Royal Society of London. A.
  Mathematical and Physical Sciences}\ }\textbf {\bibinfo {volume} {379}},\
  \bibinfo {pages} {231} (\bibinfo {year} {1982})}\BibitemShut {NoStop}%
\bibitem [{\citenamefont {Hafskjold}\ and\ \citenamefont
  {Ikeshoji}(2002)}]{Hafskjold_2002}%
  \BibitemOpen
  \bibfield  {author} {\bibinfo {author} {\bibfnamefont {B.}~\bibnamefont
  {Hafskjold}}\ and\ \bibinfo {author} {\bibfnamefont {T.}~\bibnamefont
  {Ikeshoji}},\ }\href {\doibase 10.1103/PhysRevE.66.011203} {\bibfield
  {journal} {\bibinfo  {journal} {Phys. Rev. E}\ }\textbf {\bibinfo {volume}
  {66}},\ \bibinfo {pages} {011203} (\bibinfo {year} {2002})}\BibitemShut
  {NoStop}%
\bibitem [{\citenamefont {Nairn}\ and\ \citenamefont
  {Kilpatrick}(1972)}]{Nairn_1972}%
  \BibitemOpen
  \bibfield  {author} {\bibinfo {author} {\bibfnamefont {J.~H.}\ \bibnamefont
  {Nairn}}\ and\ \bibinfo {author} {\bibfnamefont {J.~E.}\ \bibnamefont
  {Kilpatrick}},\ }\href {\doibase 10.1119/1.1986605} {\bibfield  {journal}
  {\bibinfo  {journal} {American Journal of Physics}\ }\textbf {\bibinfo
  {volume} {40}},\ \bibinfo {pages} {503} (\bibinfo {year} {1972})}\BibitemShut
  {NoStop}%
\bibitem [{\citenamefont {Krauth}(2006)}]{Krauth_06}%
  \BibitemOpen
  \bibfield  {author} {\bibinfo {author} {\bibfnamefont {W.}~\bibnamefont
  {Krauth}},\ }\href@noop {} {\emph {\bibinfo {title} {Statistical Mechanics
  Algorithms and Computations}}}\ (\bibinfo  {publisher} {Oxford University
  Press, UK},\ \bibinfo {address} {Oxford},\ \bibinfo {year}
  {2006})\BibitemShut {NoStop}%
\bibitem [{\citenamefont {White}\ \emph {et~al.}(2000)\citenamefont {White},
  \citenamefont {Gonz\'{a}lez}, \citenamefont {Rom\'{a}n},\ and\ \citenamefont
  {Velasco}}]{White_2000}%
  \BibitemOpen
  \bibfield  {author} {\bibinfo {author} {\bibfnamefont {J.~A.}\ \bibnamefont
  {White}}, \bibinfo {author} {\bibfnamefont {A.}~\bibnamefont {Gonz\'{a}lez}},
  \bibinfo {author} {\bibfnamefont {F.~L.}\ \bibnamefont {Rom\'{a}n}}, \ and\
  \bibinfo {author} {\bibfnamefont {S.}~\bibnamefont {Velasco}},\ }\href
  {\doibase 10.1103/PhysRevLett.84.1220} {\bibfield  {journal} {\bibinfo
  {journal} {Phys. Rev. Lett.}\ }\textbf {\bibinfo {volume} {84}},\ \bibinfo
  {pages} {1220} (\bibinfo {year} {2000})}\BibitemShut {NoStop}%
\bibitem [{\citenamefont {Henderson}(1983)}]{Henderson_1983}%
  \BibitemOpen
  \bibfield  {author} {\bibinfo {author} {\bibfnamefont {J.~R.}\ \bibnamefont
  {Henderson}},\ }\href {\doibase 10.1080/00268978300102661} {\bibfield
  {journal} {\bibinfo  {journal} {Molecular Physics}\ }\textbf {\bibinfo
  {volume} {50}},\ \bibinfo {pages} {741} (\bibinfo {year} {1983})}\BibitemShut
  {NoStop}%
\bibitem [{\citenamefont {Frenkel}\ and\ \citenamefont
  {Smit}(2002)}]{frenkel-smit}%
  \BibitemOpen
  \bibfield  {author} {\bibinfo {author} {\bibfnamefont {D.}~\bibnamefont
  {Frenkel}}\ and\ \bibinfo {author} {\bibfnamefont {B.}~\bibnamefont {Smit}},\
  }\href@noop {} {\emph {\bibinfo {title} {Understanding Molecular Simulation:
  From Algorithms to Applications}}}\ (\bibinfo  {publisher} {Academic Press},\
  \bibinfo {year} {2002})\BibitemShut {NoStop}%
\end{thebibliography}
%

\appendix

\section{Details of the simulation-assisted calculation\label{sec:Appendix}}

The configuration integral of the four-particle system $Z_{4}$, can
be expressed in terms of the partition functions for the systems of
fewer particles
\[
Z_{4}=6Z_{1}^{4}-12Z_{1}^{2}Z_{2}+3Z_{2}^{2}+4Z_{1}Z_{3}+\tau_{4}\:.
\]
To obtain an analytic expression for $Z_{4}$ in the low number density
branch, we have adopted an approximate expression based on the known
expressions for $Z_{2}$, $Z_{3}$, and the known terms for $\tau_{4}$
(see Eqs. (\ref{eq:biVgen}), (\ref{eq:Z234}) and Table \ref{tab:TauiComp}).
We approximate $\tau_{4}$ as its truncated form to order $R_{\textrm{eff}}^{-2}$
by assuming 
\[
\tau_{4}/24=V\, b_{4}-\, A\, a_{4}+\mathsf{J}\, c_{4,\mathsf{J}}+\mathsf{K}\, c_{4,\mathsf{K}}+R_{\textrm{eff}}^{-1}c_{4,-1}+R_{\textrm{eff}}^{-2}c_{4,-2}
\]
with unknown coefficients $c_{4,\mathsf{J}},\, c_{4,\mathsf{K}},\, c_{4,-1}$
and $c_{4,-2}$. From Eqs. (\ref{eq:FlogQ}) and (\ref{eq:PwThermo})
is obtained the analytic expression to fit $\beta P_{w}\left(R_{\textrm{eff}}\right)$.
To ensure $\Delta\tau_{4}=0$ we should restrict%
{} the fit to values $R_{\textrm{eff}}>1.5$. However, we obtain a better
description analyzing the low density behavior up to $R_{\textrm{eff}}=1.4$.
Performing this careful fit, our best estimation is $c_{4,\mathsf{J}}=-1.5162(500)$.
Once $c_{4,\mathsf{J}}$ is fixed, we analyze fitting functions with
two parameters. By fixing $c_{4,-1}=0$, we found our best fitting
coefficients $c_{4,\mathsf{K}}=-1.51495(600)$ and $c_{4,-2}=0.4181(80)$
with a regression coefficient $r^{2}=0.999999997$.

The four HS system in a spherical cavity reaches the highest number
density at a tetrahedral packing configuration. Focusing in the limiting
behavior of the wall pressure we fit
\[
\beta P_{w}A\left(R_{\textrm{eff}}-R_{\textrm{m}}\right)=\sum_{i=0}^{3}d_{4i}\left(R_{\textrm{eff}}-R_{\textrm{m}}\right)^{i}\:,
\]
where $d_{4i}$ stands for the four fitting parameters. The best
estimate for $d_{40}$ is $8.99145(80)$ which we replace by the natural
number $d_{40}=9$. Once $d_{40}$ is fixed we obtain our best estimate
for $d_{41}$ given by $-7.1745(600)$ which we represent by $d_{41}=-43/6$.
The other two fitted coefficients were $d_{42}=88.7675$ and $d_{43}=414.303$
with uncertainties of $1.0$ and $15.0$, respectively. Using the
obtained low density branch of $\beta P_{w}$ we evaluated its order
four series expansion at $R_{\textrm{eff}}=1.4$. A fitting method
was implemented that reproduce the local behavior of $\beta P_{w}$
at $R_{\textrm{eff}}=1.4$, its asymptotic properties at $R_{\textrm{eff}}\rightarrow R_{\textrm{m}}$,
and use three extra parameters to fit $\beta P_{w}$ for $R_{eff}\in(0.6124,1.4)$.
In this way we obtained an expression for $\beta P_{w}(R_{\textrm{eff}})$
for all the range of possible values of $R_{\textrm{eff}}\in(R_{\textrm{m}},+\infty)$.
The obtained function has two analytic branches that smoothly join
at $R_{\textrm{eff}}=1.4$ with a discontinuity at the fifth order
derivative.

\begin{table}[!t]
\centering{}%
\begin{tabular}{|c|c|}
\hline 
$C$ & $44840.210944988285$\tabularnewline
\hline 
$q_{2}$ & $44.38376370350302$\tabularnewline
\hline 
$q_{3}$ & $-138.10085954897494$\tabularnewline
\hline 
$q_{4}$ & $0$\tabularnewline
\hline 
$q_{5}$ & $1810.4429111889856$\tabularnewline
\hline 
$q_{6}$ & $-6779.781775837297$\tabularnewline
\hline 
$q_{7}$ & $10179.60436361462$\tabularnewline
\hline 
$q_{8}$ & $-893.2730590241717$\tabularnewline
\hline 
$q_{9}$ & $-19232.535597318078$\tabularnewline
\hline 
$q_{10}$ & $28429.173137489295$\tabularnewline
\hline 
$q_{11}$ & $-17624.014081391546$\tabularnewline
\hline 
$q_{12}$ & $4211.867673698297$\tabularnewline
\hline 
\end{tabular}\protect\caption{Coefficients for $Z_{4}$ in the high density branch that corresponds
to $R_{eff}<1.4$.\label{tab:Z4coeff}}
\end{table}
The expression of $Z_{4}$ for the high density branch, that has an
extra non-analytic point (a discontinuity) at $R_{\textrm{eff}}=1/\sqrt{2}$
(see\ref{sec:Simulation-assisted-quasi-exact-}), is found from Eqs.
(\ref{eq:FlogQ}) and (\ref{eq:DiffF}). The expression for $R_{\textrm{eff}}<1.4$
is
\[
Z_{4}=\textrm{pf}\, C\,\left(R_{\textrm{eff}}-R_{\textrm{m}}\right)^{9}\exp\left[\sum_{i=1}^{12}q_{i}\left(R_{\textrm{eff}}-R_{\textrm{m}}\right)^{i}\right]\:,
\]
with $\textrm{pf}=1$ if $R_{\textrm{eff}}\geq1/\sqrt{2}$, $\textrm{pf}=1/2$
if $R_{\textrm{eff}}<1/\sqrt{2}$ and $q_{1}=-43/6$. The remaining
coefficients are given in Table \ref{tab:Z4coeff}.
\end{document}